\documentclass{emulateapj}

\begin{document}

\shortauthors{Luhman}

\shorttitle{Census of 32 Ori}

\title{A Census of the 32 Ori Association with Gaia\altaffilmark{1}}

\author{K. L. Luhman\altaffilmark{2,3}}

\altaffiltext{1}
{Based on observations made with the Gaia mission, the Two Micron All
Sky Survey, the Wide-field Infrared Survey Explorer, the LAMOST survey,
the Sloan Digital Sky Survey IV, the NASA Infrared Telescope Facility, 
and Cerro Tololo Inter-American Observatory,}

\altaffiltext{2}{Department of Astronomy and Astrophysics,
The Pennsylvania State University, University Park, PA 16802, USA;
kll207@psu.edu}

\altaffiltext{3}{Center for Exoplanets and Habitable Worlds, The
Pennsylvania State University, University Park, PA 16802, USA}

\begin{abstract}

I have used high-precision photometry and astrometry from 
the third data release of Gaia (DR3) to identify
candidate members of the 32~Ori association.
Spectral types and radial velocities have been measured for
subsets of the candidates using new and archival spectra.
For the candidates that have radial velocity measurements,
I have used $UVW$ velocities to further constrain their membership,
arriving at a final catalog of 169 candidates.
I estimate that the completeness of the survey is $\sim90$\%
for spectral types of $\lesssim$M7 ($\gtrsim0.06$~$M_\odot$).
The histogram of spectral types for the 32~Ori candidates exhibits a 
maximum at M5 ($\sim0.15$~$M_\odot$), resembling the distributions
measured for other young clusters and associations in the solar neighborhood.
The available $UVW$ velocities indicate that the association is expanding,
but they do not produce a well-defined kinematic age.
Based on their sequences of low-mass stars in color-magnitude diagrams,
the 32~Ori association and Upper Centaurus-Lupus/Lower
Centaurus-Crux (UCL/LCC) are coeval to within $\pm1.2$~Myr, and they
are younger than the $\beta$~Pic moving group by $\sim3$~Myr, which agrees with
results from previous analysis based on the second data release of Gaia.
Finally, I have used mid-infrared (IR) photometry from the Wide-field 
Infrared Survey Explorer to check for excess emission from
circumstellar disks among the 32~Ori candidates.  Disks are detected for 18 
candidates, half of which are reported for the first time in this work. 
The fraction of candidates at $\leq$M6 that have full, transitional, or 
evolved disks is 10/149=$0.07^{+0.03}_{-0.02}$, which is consistent with the 
value for UCL/LCC.

\end{abstract}

\section{Introduction}
\label{sec:intro}

The identification of members of nearby young associations ($\lesssim$100~pc,
$\lesssim100$~Myr) is important for studies of the formation and early
evolution of stars and planets \citep{zuc04,tor08}.
Surveys for association members have utilized multiple approaches,
which have consisted primarily of the following:
(1) selecting candidates for young nearby stars from all-sky surveys in X-rays
or UV emission
\citep{kas97,mam99,web99,tor00,zuc01a,shk09,shk12,rod11,rod13,kas17,bow19,bin20},
confirming their youth with spectroscopy, and identifying the associations to
which they might belong with available kinematic data;
(2) selecting candidate members of specific associations via
photometry and proper motions from wide-field imaging surveys
\citep{son04,mal13,gag14a,ell16,rie17b,shk17,sch19}, sometimes in conjunction
with X-ray and UV emission \citep{schl10,schl12a,schl12c,bin15,bin16,bin18}, and
confirming their youth with spectroscopy; 
(3) selecting candidates for late-type dwarfs with photometry from
wide-field infrared (IR) surveys, confirming their cool nature via
spectroscopy, and obtaining kinematic data for objects with spectral 
evidence of youth to assess membership in young associations
\citep{ric10,del12,liu13,liu16,kel15,sch14,sch16,bur16,bes17,bar18,gag18a};
(4) selecting candidates for late-type members of associations via IR
photometry and proper motions from wide-field surveys and confirming
their youth and cool nature with spectroscopy
\citep{gag14a,gag14b,gag14c,gag15a,gag15b,gag15c,gag17,gag18b,all16}.
Parallax measurements from the Hipparcos mission also have been utilized
in the identification of the brightest members of associations
\citep{zuc00,zuc01c}. These surveys have produced hundreds of candidate 
members of nearby associations, although it is likely that there remains
significant incompleteness, particularly among low-mass stars and brown dwarfs.

Progress in cataloging the members of nearby associations has been driven
by advances in wide-field photometric and astrometric surveys. 
That progress has been especially rapid in recent years due to the Gaia mission
\citep{per01,deb12,gaia16b}, which is performing an all-sky survey to measure
high-precision photometry, proper motions, and parallaxes for more than a
billion stars as faint as $G\sim20$.
Gaia data have been used to identify both new associations in the solar 
neighborhood \citep{oh17,fah18,gag18e,kou19,mei19} and new candidate members 
of known associations \citep{gag18c,gag18d,lee19,zuc19,ujj20}.

In this paper, I present a survey for members of the association
containing the B5 star 32~Ori \citep{mam07,bur16,bel17} using 
the third data release of Gaia \citep[DR3,][]{bro21,val22}. 
I have used astrometry and photometry from Gaia DR3 to identify candidate 
members of 32~Ori (Section~\ref{sec:ident}) and have performed spectroscopy on 
a subset of those candidates to measure their spectral types, diagnostics of 
youth, and radial velocities (Section~\ref{sec:spec}). 
For the candidates with measured radial velocities, membership is further
constrained using $UVW$ velocities (Section~\ref{sec:uvw}). 
I compare my final catalog of candidates to previous membership studies 
of 32~Ori (Section~\ref{sec:prev}) and I use it to
study the association in terms of its initial mass function (IMF), kinematic
and isochronal ages, and circumstellar disks (Section~\ref{sec:pop}).

\section{Identification of Candidate Members}
\label{sec:ident}

\subsection{Kinematic Selection Criteria}
\label{sec:kin}

The analysis in this study was initially based on the second data release
of Gaia (DR2) and was updated with Gaia DR3 when it became available.
The Gaia data employed in my analysis consist of the following:
photometry in bands at 3300--10500~\AA\ ($G$), 3300--6800~\AA\ ($G_{\rm BP}$),
and 6300-10500~\AA\ ($G_{\rm RP}$); proper motions and parallaxes
($G\lesssim20$); radial velocities ($G\lesssim14$);
and the renormalized unit weight error \citep[RUWE,][]{lin18}.
The latter provides a measure of the goodness of fit for the astrometry.
As in my previous work with Gaia data \citep[e.g.,][]{luh20u,luh22sc}, 
I adopt a threshold of RUWE$<$1.6 when selecting astrometry that is likely 
to be reliable.

Since radial velocity data are available for only a small fraction
of the stars from Gaia DR3 that have proper motion and parallax measurements,
I rely on the latter parameters for the initial kinematic selection
of candidate members of 32~Ori.  To reduce projection effects, I analyze 
the Gaia astrometry in terms of a ``proper motion offset"
($\Delta\mu_{\alpha,\delta}$), which is defined as the difference between the 
observed proper motion of a star and the motion expected at the celestial 
coordinates and parallactic distance of the star for a specified 
space velocity \citep[e.g.,][]{esp17,luh20u,luh22sc}.
In this study, the proper motion offsets are calculated relative to the 
motions expected for a space velocity of $U, V, W = -12, -19, -9$~km~s$^{-1}$,
which approximates the median velocity of 32~Ori members 
\citep[][Section~\ref{sec:uvw}]{bel17}. For parallactic distances, I adopt 
the geometric values estimated by \citet{bai21} from DR3 parallaxes.

\begin{figure}
\epsscale{1}
\plotone{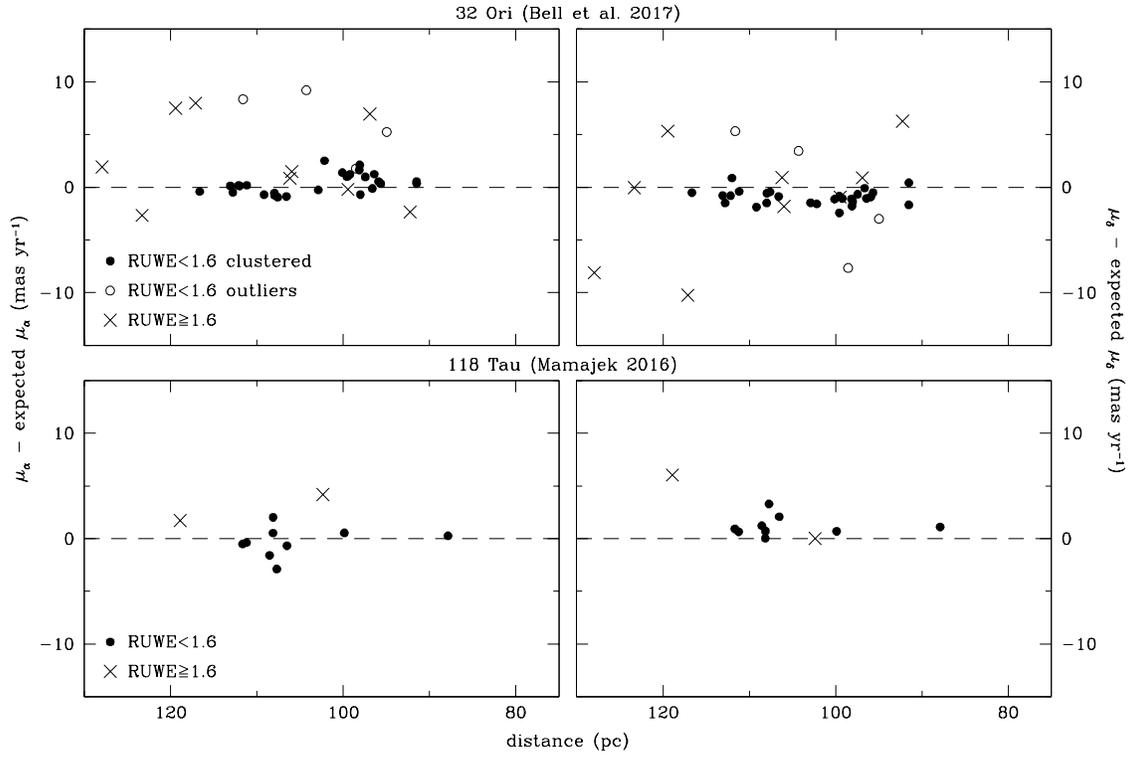}
\caption{
Proper motion offsets versus parallactic distance for previously proposed
members of the 32~Ori and 118~Tau associations \citep{mam16,bel17}.
}
\label{fig:pp1}
\end{figure}

To develop kinematic criteria for selecting candidate members of 32~Ori from
Gaia DR3, I began by considering the 46 proposed members from \citet{bel17}.
Parallax measurements are available from Gaia DR3 for 44 of those 46 sources.
The objects that lack parallaxes consist of 2MASS J05235565+1101027 
\citep[M5,][]{bel17} and WISE J052857.68+090104.4 \citep[L1,][]{bur16}; 
the former is in Gaia DR3 while the latter was not detected by Gaia.
The candidates with parallax data are plotted in diagrams of proper motion
offsets versus parallactic distance in the top row of Figure~\ref{fig:pp1}.
Among the 32 stars with RUWE$<$1.6, 27 are well-clustered in those parameters
and are plotted with one symbol while the remaining five stars are outliers
and are plotted with a second symbol. Stars with RUWE$\geq$1.6 are shown
with a third symbol; three are well-clustered and nine are outliers (three
are beyond the limits of the diagrams). 
The five outliers with RUWE$<$1.6 consist of 32~Ori~A, THOR~9A, THOR~26,
THOR~16, and THOR~34. The latter three also appear below the sequence
formed by the well-clustered stars in color-magnitude diagrams (CMDs)
constructed from Gaia photometry (Section~\ref{sec:phot}), so they are 
classified as non-members in this work. 
Although they have low RUWE, the astrometry of 32~Ori~A and
THOR~9A could have errors that are caused by their companions,
so they are retained as candidate members.

During the application of preliminary kinematic criteria based on the
32~Ori members from \citet{bel17}, I realized that the small group
associated with the B8 star 118~Tau from \citet{mam16} is spatially adjacent to 
the 32~Ori sample and shares a roughly similar space velocity, indicating that 
the stars near 32~Ori and 118~Tau may be members of the same association.
Therefore, I also considered the 11 candidate members of the
118~Tau group from \citet{mam16} when designing the final selection criteria.
The proper motion offsets and distances of those stars are shown in the bottom
row of Figure~\ref{fig:pp1}, where stars with RUWE$<$1.6 and RUWE$\geq$1.6
are plotted with different symbols. The nine stars with RUWE$<$1.6 have similar
offsets and distances as the 32~Ori candidates. For my survey, I have adopted
$\mid\Delta\mu_{\alpha,\delta}\mid<4$~mas~yr$^{-1}$ as selection
criteria for new candidates, which encompass all of the well-clustered 32~Ori
and 118~Tau candidates in Figure~\ref{fig:pp1}.
I applied those $\Delta\mu_{\alpha,\delta}$ criteria and the
photometric criteria from Section~\ref{sec:phot} to the Gaia DR3 catalog
for ranges of equatorial coordinates and distance that are large enough to
encompass the well-clustered 32~Ori and 118~Tau candidates in 
Figure~\ref{fig:pp1}, and I expanded those ranges until few additional
candidates were identified by additional expansion. Based on that analysis, 
I have adopted 50--90$\arcdeg$, 0--45$\arcdeg$, and 70--120~pc as selection
criteria for right ascension, declination, and distance, respectively.
In the left panel of Figure~\ref{fig:map}, the previously proposed members of
118~Tau and 32~Ori are plotted on a map of equatorial coordinates for that
survey field. In addition, I have required that candidates have RUWE$<$1.6 and
$\sigma_{\pi}<1$~mas.
I have retrieved from Gaia DR3 all sources that satisfy the preceding
selection criteria, resulting in a sample of 244 candidates after 
the stars from \citet{mam16} and \citet{bel17} are excluded.

In Section~\ref{sec:uvw}, analysis of the space velocities of candidates 
from this survey demonstrates that the previous samples
associated with 118~Tau and 32~Ori do indeed reside within 
a single association. Since 32~Ori has an earlier spectral type than
118~Tau, I have adopted the former as the name for the association.
For the remainder of this study, I refer to the 118~Tau group only when
specifying the sample of stars from \citet{mam16}.

\subsection{Photometric Selection Criteria}
\label{sec:phot}

\begin{figure}
\epsscale{1.1}
\plotone{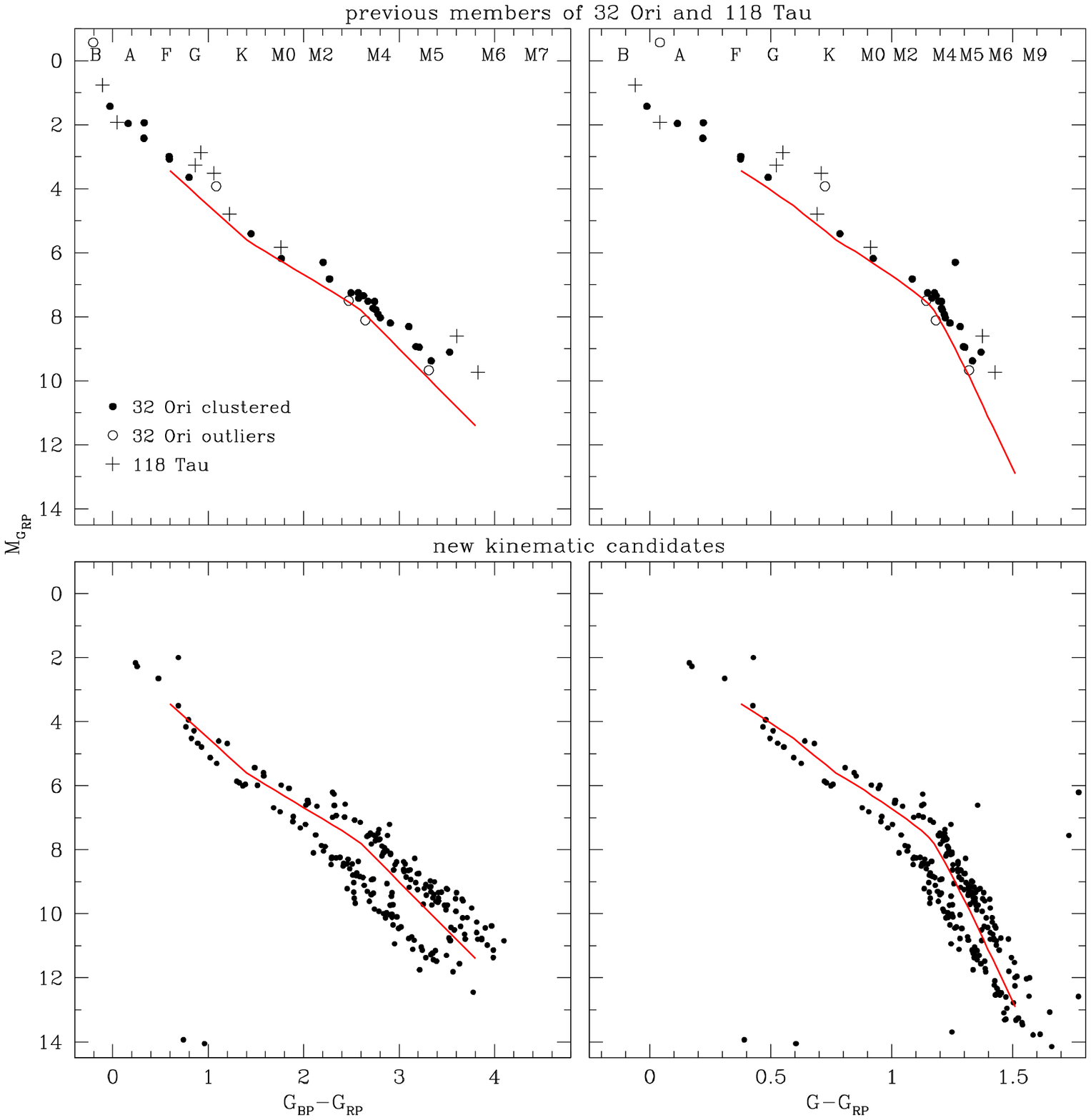}
\caption{
Top: $M_{G_{\rm RP}}$ versus $G_{\rm BP}-G_{\rm RP}$ and $G-G_{\rm RP}$
for previously proposed members of 32~Ori and 118~Tau that
have RUWE$<$1.6 \citep{mam16,bel17}. The boundaries used for selecting
candidate members of Sco-Cen by \citet{luh22sc} are marked (red lines).
Bottom: New kinematic candidate members 
that have RUWE$<$1.6 and $\sigma_{\pi}<1$~mas (Section~\ref{sec:kin}).
For reference, the spectral types that correspond to the colors of young
stars are indicated \citep{luh22sc}.
}
\label{fig:cmd1}
\end{figure}

The candidate members of the 32 Ori association selected via kinematics in
the previous section can be further refined with CMDs.
In the top row of Figure~\ref{fig:cmd1}, I present CMDs of
$M_{G_{\rm RP}}$ versus $G_{\rm BP}-G_{\rm RP}$ and $G-G_{\rm RP}$ for the 
previously proposed members of 118~Tau and 32~Ori that have RUWE$<$1.6,
excluding photometry with errors greater than 0.1~mag.
The 32~Ori stars are plotted with symbols based on whether they are
outliers in Figure~\ref{fig:pp1}. As mentioned in the previous section,
three of the 32~Ori kinematic outliers appear below the sequences formed by
the other members, and hence are rejected as non-members.
Although the number of 118~Tau candidates is small, their positions
in the CMDs are consistent with the same age as the 32~Ori candidates
if they consist of a mixture of single stars and unresolved binaries.
Through analysis of Gaia DR2 CMDs, \citet{luh20u} found that the 32~Ori
members from \citet{bel17} have a similar age as Upper Centaurus-Lupus/Lower
Centaurus-Crux (UCL/LCC), which is one of the oldest populations
in the Scorpius-Centaurus (Sco-Cen) OB association ($\sim$20~Myr).
In Figure~\ref{fig:cmd1}, I have included the boundaries that 
\citet{luh22sc} used for selecting photometric candidates in UCL/LCC and
other populations in Sco-Cen. Those boundaries closely follow the lower
envelopes of the sequences of previously proposed members of 32~Ori and
118~Tau, so they are used for selecting candidates in this survey as well.

The bottom row of Figure~\ref{fig:cmd1} shows CMDs for the kinematic
candidates selected in the previous section. In each CMD, the candidates form
two distinct populations that consist of stars near the main sequence and
a brighter band of stars that aligns with the sequence of previously
proposed members of 32~Ori and 118~Tau. 
A few stars appear within the sequence of young stars in the
$G_{\rm BP}-G_{\rm RP}$ CMD but are much redder than that sequence
in the $G-G_{\rm RP}$ CMD, which is likely a reflection of erroneous
photometry in $G$ due to contamination from a close companion \citep{eva18}.
I have rejected kinematic candidates
that appear below the boundary from \citet{luh22sc} in either CMD unless
they exhibit IR excess emission in photometry from 
the Wide-field Infrared Survey Explorer \citep[WISE,][]{wri10}. 
Stars with IR excesses are retained because an excess may indicate
the presence of a disk, and a disk-bearing star can appear underluminous
in a CMD if the disk is edge-on or if accretion-related emission is
present at shorter optical wavelengths. Two candidates appear below the
CMD boundaries and exhibit IR excesses, consisting of Gaia DR3 
3415666239289777664 and 3308700559817832576 \citep[LDS~5606~A,][]{rod14}.
I also have rejected a few additional candidates that appear below the 
32~Ori sequence in other CMDs such as 
$M_{G_{\rm RP}}$ versus $G_{\rm RP}-K_s$ where $K_s$ is from
the Point Source Catalog of the Two Micron All Sky Survey
\citep[2MASS,][]{skr03,skr06}. These photometric criteria are satisfied 
by 122 of the 244 kinematic candidates.

\begin{figure}
\epsscale{1}
\plotone{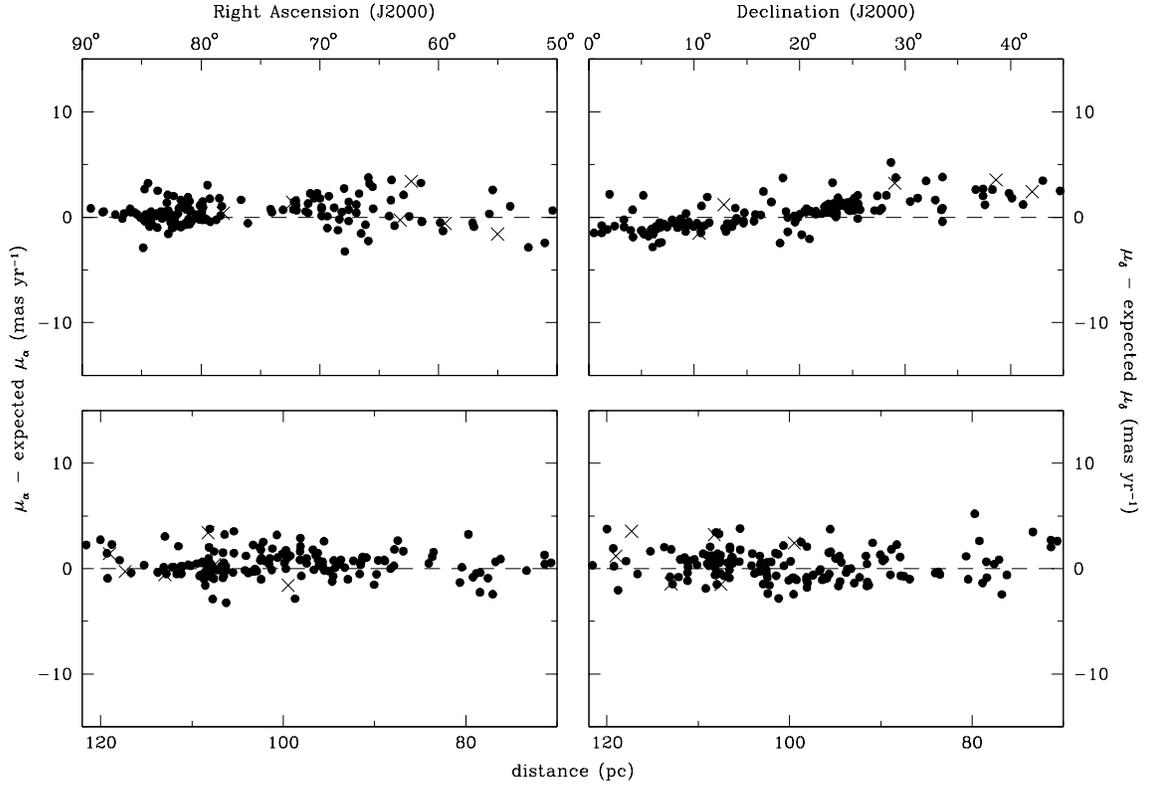}
\caption{
Proper motion offsets versus celestial coordinates (top) and parallactic 
distance (bottom) for candidate members of the 32 Ori association 
based on kinematics and CMDs (Section~\ref{sec:phot}).
Six of the candidates are rejected via $UVW$ velocities and other data
(crosses, Section~\ref{sec:uvw}).}
\label{fig:pp2}
\end{figure}

When I combine the 122 candidates with the 36 previously
proposed members of 118~Tau and 32~Ori that have RUWE$<$1.6 and are
well-clustered in Figure~\ref{fig:pp1} (Section~\ref{sec:kin}), 
I arrive at a sample of 158 candidate members. In Figure~\ref{fig:pp2}, 
I have plotted that sample in diagrams of $\Delta\mu_{\alpha}$ versus right 
ascension and distance and $\Delta\mu_{\delta}$ versus declination
and distance.
The candidates exhibit a correlation between $\Delta\mu_{\delta}$ and 
declination, which is suggestive of expansion in that direction.
In Sections~\ref{sec:uvw} and \ref{sec:uvw2}, evidence of expansion is 
investigated in more detail using the $UVW$ velocities for the candidates 
that have radial velocity measurements.

\subsection{Additional Candidates}
\label{sec:add}

In the previous section, I identified 158 stars that satisfy kinematic and
photometric criteria for membership in 32~Ori. In this section, I describe 
17 additional stars that are included in my catalog of candidates.

As discussed in Section~\ref{sec:kin}, two of the proposed members of 32~Ori
from \citet{bel17}, 32~Ori~A and THOR~9A, are kinematic outliers in
Figure~\ref{fig:pp1} but are retained as candidates because their astrometry
may be affected by companions.

THOR~10, THOR~32, THOR~8A, HHJ~430, Gaia DR3 3284091290564991104,
and Gaia DR3 3428203622488886784 satisfy the 
kinematic and photometric criteria for membership but have RUWE$>$1.6. 
For each star, the proper motion and parallax measurements are similar 
between DR2 and DR3 (and the first data release for THOR~10),
which suggests that their astrometry may be reliable. 
The proper motion and parallax of THOR~8A are also similar to
those of its candidate companion, THOR~8B, which is among the candidate members
from the previous section.
In addition, the six stars exhibit spectroscopic evidence of youth
\citep[][this work]{opp97,bel17,sta20}. Therefore, I have adopted them
as candidate members.

I searched for companions to the candidates from the previous section
that are in Gaia DR3 but did not satisfy all of the selection criteria.
To do that, I retrieved sources from DR3 that are located within $5\arcsec$
from the candidates and that (1) fail the kinematic criteria but satisfy
the photometric criteria and share roughly similar parallaxes and proper
motions as their neighboring candidates 
or (2) satisfy the kinematic criteria but lack the photometry for the CMDs.
The resulting candidate companions consist of 118~Tau~Ba, THOR~9B, 
THOR~14Ab, HD~245924~B, RXJ0416.0+3325~B, and sources 241638674007806336, 
3416960501912034816, 3237714611659033728, and 44866400900149248 from Gaia DR3.

\citet{bur16} identified the young L dwarf WISE J052857.68+090104.4
\citep{tho13} as a possible member of 32~Ori based on its proper motion, 
radial velocity, and a rough estimate of its distance that was derived from
photometry.  Since it does not appear in Gaia DR3 and lacks a parallax 
measurement, I am unable to assess its membership with the same level of
scrutiny as the candidates selected in this work.  As a result, it has 
been omitted from my census of 32~Ori.

\subsection{Completeness and Contamination}
\label{sec:comp}

The source catalog from Gaia DR3 has a high level of completeness at
$G\lesssim19$--20 for most of the sky \citep{bou20,fab21}, which
corresponds to spectral types earlier than M7--M8 in 32~Ori.
I have calculated the fraction of DR3 sources in my survey field
that satisfy the following criteria utilized in my selection of candidates:
$\sigma_{\pi}<1$~mas, RUWE$<$1.6, and $\sigma_{BP}<0.1/\sigma_{RP}<0.1$
or $\sigma_G<0.1/\sigma_{RP}<0.1$. The resulting fraction varies from
$\sim80$\% at $G=5$ to $\sim95$\% at $G=19$ and quickly declines at
$G\gtrsim$19.5. Thus, my census of 32~Ori should be $\sim90$\% complete
for spectral types of $\lesssim$M7 and for the ranges of celestial coordinates
and distance that have been considered.

I have estimated the field star contamination among the candidate
members of 32~Ori by repeating the kinematic and photometric selection
procedures with the $\Delta\mu_{\alpha,\delta}$ criteria shifted
by (+10,+10), (+10,0), (+10,$-$10), (0,+10), (0,$-$10), ($-$10,+10), ($-$10,0),
and ($-$10,$-$10)~mas~yr$^{-1}$. On average, these eight samples contain
$\sim5$ A/F/G stars, $\sim4$ M3--M6 stars, and $\sim5$ stars later than M6,
which correspond to $\sim9$\% of the 158 stars that satisfy kinematic 
and photometric criteria for membership in 32~Ori. 
These results are roughly consistent with the number of candidates that are
rejected via their $UVW$ velocities in Section~\ref{sec:uvw}.
Those velocities are available for 69\% of the 32~Ori candidates
and lead to the rejection of three F/G/K stars and three M stars.

\section{Spectroscopy of Candidates}
\label{sec:spec}

I have used new and archival spectra of some of the candidate members
of 32~Ori from Section~\ref{sec:ident} to measure their spectral types and 
radial velocities. 
The telescopes, instruments, and observing modes are summarized in 
Table~\ref{tab:log} and the measured classifications and velocities
are provided in Table~\ref{tab:spec}.

\subsection{Spectral Classifications}

To measure spectral types, I obtained IR spectra with SpeX \citep{ray03}
at the NASA Infrared Telescope Facility (IRTF) on the nights of 2020
January 1, 20, and 21 and I collected optical spectra with the
Cerro Tololo Ohio State Multi-Object Spectrograph (COSMOS)\footnote{COSMOS 
is based on an instrument described by \citet{mar11}.} at the 4~m Blanco 
telescope at the Cerro Tololo Inter-American Observatory (CTIO) on the night of 
2021 January 2. In addition, I made use of the following archival data: 
an optical spectrum taken with the RC spectrograph at the CTIO 1.5~m telescope 
on the night of 2011 December 18 through program UR11b-10 (E. Mamajek), 
an IR spectrum taken with the IRTF/Spex on the night of 2017 December 26 
through program 2017B083 (Z. Zhang), and optical spectra from the seventh data 
release of the Large Sky Area Multi-Object Fiber Spectroscopic Telescope survey
\citep[LAMOST;][]{cui12,zha12}.

\begin{figure}
\epsscale{1}
\plotone{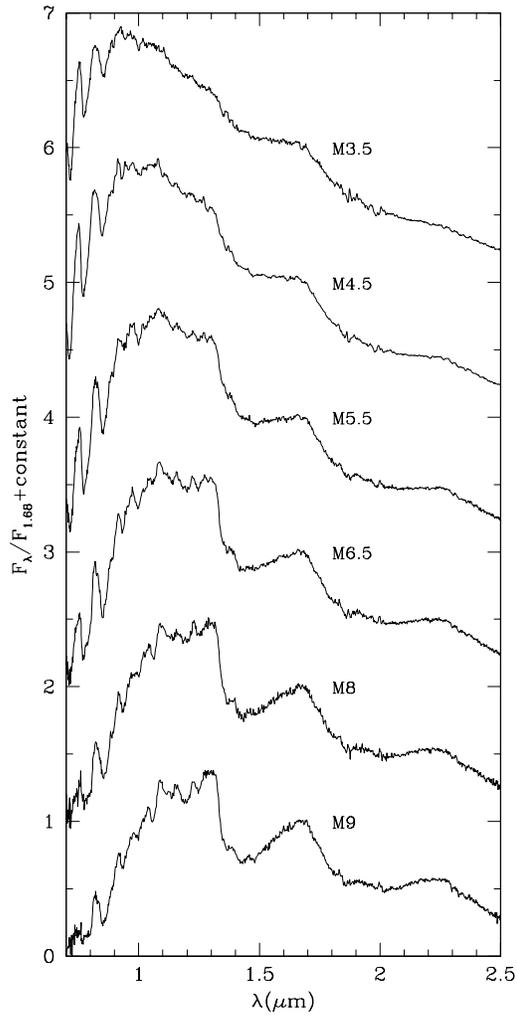}
\caption{
Examples of spectra of candidate members of the 32~Ori association, which are 
displayed at a resolution of $R=150$. The data used to create this figure
are available.
}
\label{fig:ir}
\end{figure}

The IRTF/SpeX data were reduced with the Spextool package \citep{cus04}, 
which included correction of telluric absorption \citep{vac03}.
The optical spectra from CTIO were reduced with routines within IRAF.
Examples of the reduced IR spectra are presented in Figure~\ref{fig:ir}.
The reduced optical and IR spectra are available in an electronic file
associated with Figure~\ref{fig:ir}.

Spectral classifications were performed for 96 objects, some of which
were classified with data from multiple spectrographs.
I assessed the ages of the targets using Li absorption at 6707~\AA\ and
gravity-sensitive features like the Na doublet near 8190~\AA\ and the
near-IR H$_2$O absorption bands. The resolution of the LAMOST data is too low 
for the reliable measurement of weak Li absorption ($W_\lambda\lesssim0.3$~\AA).
For all targets, the spectra are consistent with the age of 32~Ori,
although the available data are insufficient to distinguish between
young stars and field dwarfs for spectral types near M2--M3, where Li is
expected to be depleted in 32~Ori members \citep{bel17} and the 
available gravity-sensitive features exhibit only subtle changes with age.
Most of those stars do have measurements of radial velocities that support
their membership (Section~\ref{sec:uvw}).
Spectral types were measured from the optical spectra through comparison
to field dwarf standards for $<$M5 \citep{hen94,kir91,kir97} and
averages of dwarf and giant standards for $\geq$M5 \citep{luh97,luh99}. 
The IR spectra were classified with standard spectra derived from 
optically-classified young sources \citep{luh17}.
Among the 96 sources for which spectral types were measured, 62 have
been classified for the first time in this work.
Spectral classifications remain unavailable for 26 of the candidates
from Section~\ref{sec:ident}, six of which have separations of 
$\lesssim1\arcsec$ from other stars.

\subsection{Radial Velocities}

I obtained high-resolution IR spectra of 39 candidate members of 32~Ori using 
iSHELL at the IRTF \citep{ray22}. 
The selected observing mode for iSHELL (Table~\ref{tab:log}) produced spectra
in orders 238--211, which span wavelengths from 2.17---2.46~\micron.
At least five exposures were collected for each target, most of which
had exposure times of 3--5~min.
The spectra were reduced using a version of Spextool (5.0.3) that was modified
for use with iSHELL. The data were corrected for telluric absorption
using the beta version of the IDL routine 
{\tt xtellcor\_model}\footnote{\url{http://irtfweb.ifa.hawaii.edu/research/dr\_resources/}}.

To measure radial velocities from the iSHELL data, I employed orders 226--217 
(2.285--2.391~\micron) because they encompassed a large number of
strong photospheric absorption lines and relatively few saturated
telluric lines. For each target, a template for its photospheric spectrum
was derived from the iSHELL data with the python package {\tt pychell} 
\citep{cal19}, where the template was initialized with a BT-Settl model 
spectrum near the temperature and surface gravity of the target 
\citep{all12,bar15}. For each order, a radial velocity was calculated
by combining the barycentric correction \citep{wri14} and the velocity shifts 
produced by cross correlating the following pairs of spectra: 
the BT-Settl and the template, the template and the telluric-corrected 
target spectrum, and the uncorrected target spectrum and the model of 
telluric absorption utilized by {\tt xtellcor\_model} \citep{vil18}. 
By including the latter cross correlation, the telluric lines served as the
source of the velocity calibration.
The cross correlations were performed with {\tt fxcor} within IRAF 
\citep{ton79}. The mean and standard deviation of the radial velocities from 
the 10 orders for each target are presented in Table~\ref{tab:spec}.
The median of the standard deviations is 1.0~km~s$^{-1}$.

\section{Refining Candidates with $UVW$ Velocities}
\label{sec:uvw}

For the candidate members of 32~Ori that have radial velocity measurements,
their membership can be further constrained with $UVW$ velocities.
I have compiled radial velocities that have errors less than 4~km~s$^{-1}$
from this work and previous studies for 113 of the 175 candidates from 
Section~\ref{sec:ident}. As done by \citet{bel17}, I have excluded the
available measurement for HD~35656 because it may be unreliable.
Multiple velocities spanning a large range have been measured for
HD~28693 \citep[][Gaia DR3]{jon20,abd22}, THOR~31 \citep[][Gaia DR3]{bel17},
and RX~J0437.4+1851~B \citep[][Gaia DR2]{wic00}, so I have not adopted
velocities for them. These stars may be spectroscopic binaries, in which
case their system velocities will need to be characterized for the kind
of analysis in this section. Some of the adopted
radial velocities are from the APOGEE-2 program within the Sloan Digital
Sky Survey IV \citep[SDSS-IV,][]{bla17,maj17,abd22}. 
The errors in those velocities are likely underestimated \citep{cot14,tsa22}.
For each of two stars (V924~Tau~A and B), I have adopted the median of the
radial velocities measured at multiple epochs by the LAMOST Medium-resolution
Survey as calibrated by \citet{zha21} and I have adopted the
standard deviation of the multiple measurements as the error.

I have used the compiled radial velocities in conjunction
with proper motions from Gaia DR3 and parallactic distances based on DR3
parallaxes \citep{bai21} to calculate $UVW$ velocities \citep{joh87}.
The velocity errors were estimated in the manner described by \citet{luh20u}.
For most stars, the errors in $U$ are larger than the errors in the other two
velocity components because they are determined primarily by the radial
velocity errors, which are usually larger than the equivalent errors
in proper motion.

\begin{figure}
\epsscale{1}
\plotone{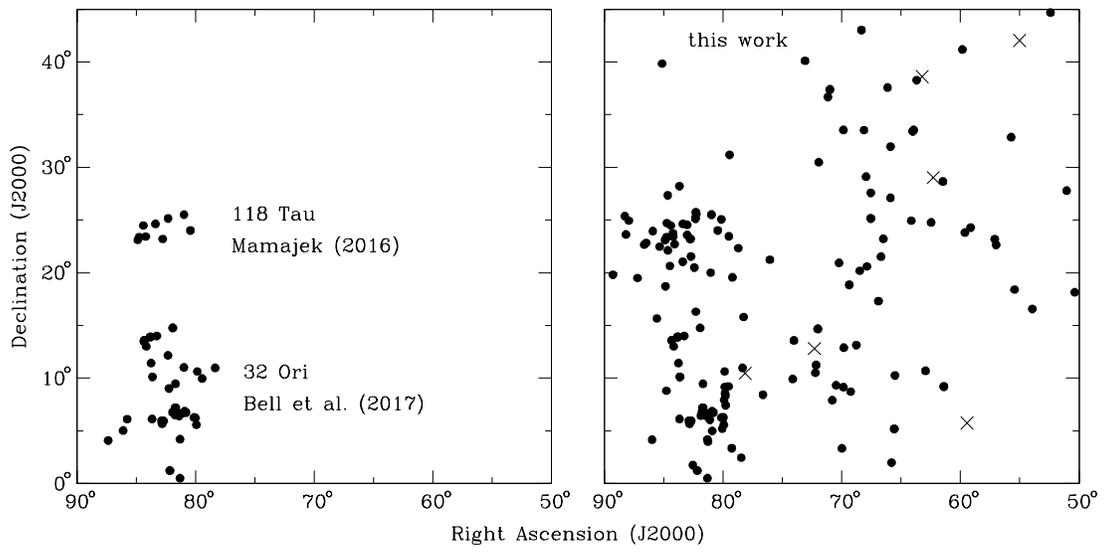}
\caption{Maps of previously proposed members of the 32~Ori and 
118~Tau associations \citep[left,][]{mam16,bel17} and the 
candidate members compiled in this work (right, Section~\ref{sec:ident}).
Six of the latter candidates are rejected via $UVW$ velocities and other data
(crosses, Section~\ref{sec:uvw}).}
\label{fig:map}
\end{figure}

\begin{figure}
\epsscale{1.1}
\plotone{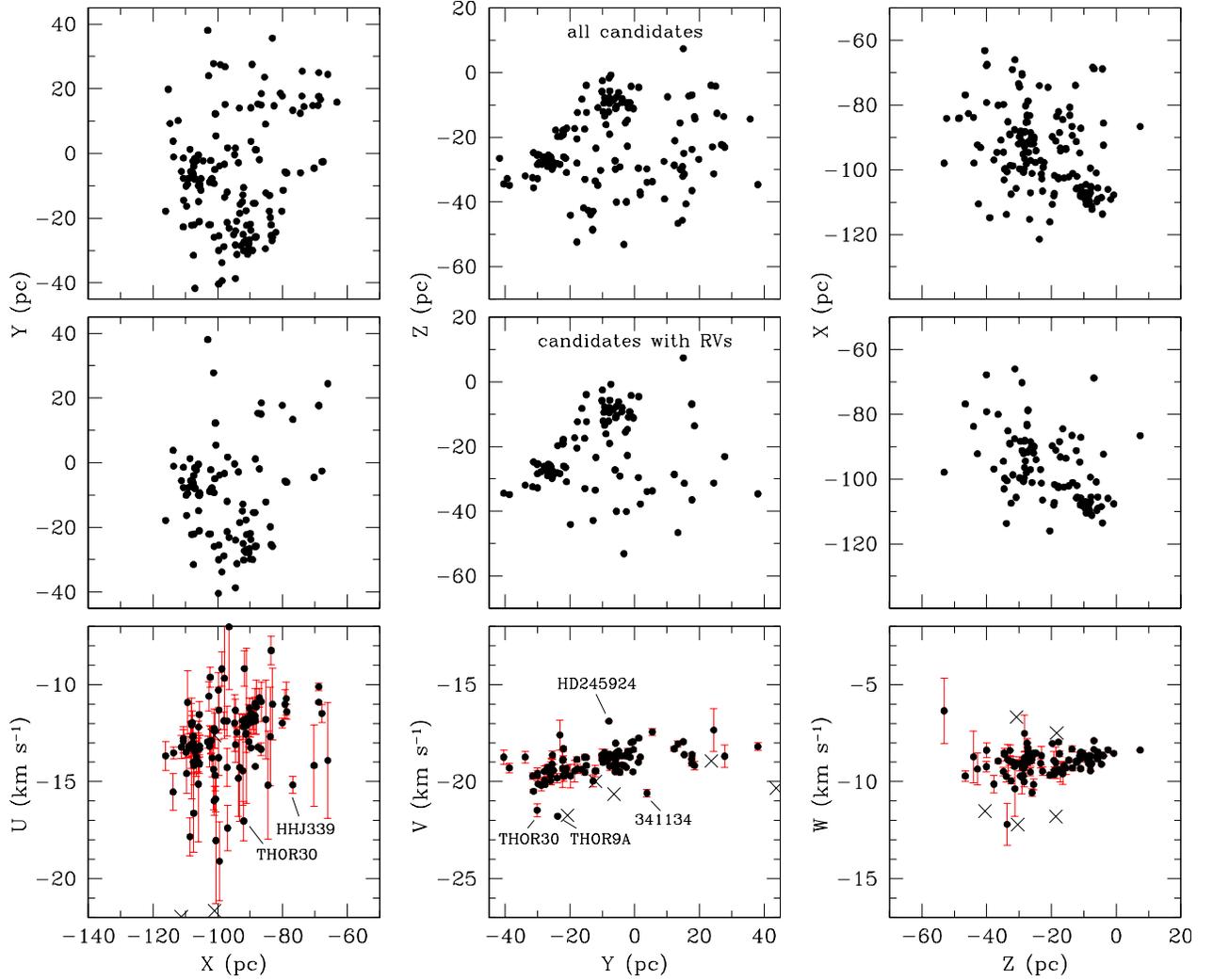}
\caption{
Galactic Cartesian coordinates for candidate members of the 32~Ori
association (top).
Candidates that have measurements of radial velocities are plotted in diagrams
of $XYZ$ and $UVW$ (middle and bottom). Six of the candidates are rejected 
via their velocities in the bottom diagrams and other data
(crosses, Section~\ref{sec:uvw}).
Modestly discrepant measurements of $UVW$ among the remaining candidates
are labeled with the source names.}
\label{fig:uvw}
\end{figure}

Before analyzing the available $UVW$ velocities for the 32 Ori candidates,
I examine the spatial distribution of the full sample of 175 candidates from
Section~\ref{sec:ident}.  I have plotted the $XYZ$ positions in Galactic 
Cartesian coordinates for all candidates in the top row of Figure~\ref{fig:uvw}.
In addition, the candidates are shown on a map of equatorial coordinates
in the right panel of Figure~\ref{fig:map}.
In those diagrams, the candidates exhibit two subclusters and a sparser
distribution of stars that extends primarily to higher values of $Y$.
As shown in Figure~\ref{fig:map}, the 32~Ori candidates from \citet{bel17}
are concentrated near one subcluster while the 118~Tau candidates from
\citet{mam16} are located in the other subcluster.
The $XYZ$ positions of the candidates that have measurements of radial
velocities (and hence $UVW$ velocities) are plotted in the middle row of
Figure~\ref{fig:uvw}. Those candidates provide a good sampling of the
full list of candidates.

In the bottom row of Figure~\ref{fig:uvw}, the available measurements of
$U$, $V$, and $W$ are plotted versus $X$, $Y$, and $Z$, respectively.
Most of the velocities are tightly clustered in these diagrams, and
each velocity component exhibits a positive correlation with
spatial position except for $W$ at $Z\lesssim-20$~pc, which is roughly flat. 
These characteristics indicate that most of the candidates
are members of a single association that is expanding.

I have rejected six candidates based on their discrepant velocities.
Their names and velocities are provided in Table~\ref{tab:reject}.
These rejected stars are marked with crosses in Figure~\ref{fig:uvw}.
Some of the measurements are beyond the limits of those diagrams.
The velocity of one star, HD~281691, is only modestly discrepant,
but it and a $7\arcsec$ companion both appear slightly below the sequence
for 32~Ori in CMDs, which further indicates that they are probably not members.
Although HD~281691 was selected as a candidate member of 32~Ori in
Section~\ref{sec:ident}, its companion was not among the candidates
because it did not satisfy the kinematic selection criteria
(the two stars straddle the thresholds).


Five additional candidates have $UVW$ velocities that are modestly discrepant,
consisting of THOR~30, THOR~9A, HD245924~A, Gaia DR3 3411342134934571520,
and HHJ~339. The discrepant measurements are labeled with the
source names in Figure~\ref{fig:uvw}. Only the first six digits of
the Gaia designation are indicated.
Two of the candidates, THOR~9A and HD245924~A, have subarcsecond companions
that are resolved by Gaia, so their discrepant measurements could be due to
astrometric errors caused by their companions. These five stars are retained 
in my catalog of candidates, but I consider their membership to be tentative.

For the 107 candidates that have measurements of $UVW$ velocities and that
are retained in my final catalog, the median velocity is
$U, V, W = -12.9, -18.9, -8.9$~km~s$^{-1}$, which is similar to the value
from \citet{bel17}.

Among the 175 candidates from Section~\ref{sec:ident}, six have been rejected
in this section. The remaining 169 candidates are presented in 
Table~\ref{tab:mem}, which includes source names from Gaia DR3, the 2MASS Point 
Source Catalog, the AllWISE Source Catalog \citep{cut13a,wri13}, and previous 
studies; equatorial coordinates, proper motion, parallax, RUWE, and photometric
magnitudes from Gaia DR3; measurements of spectral types and the
type adopted in this work; distance estimate based on Gaia DR3 parallax 
\citep{bai21}; the most accurate available radial velocity measurement
that has an error less than 4~km~s$^{-1}$; the $UVW$ velocities calculated in
this section; photometry from 2MASS and WISE; and flags 
indicating whether excesses are detected in three WISE bands and a disk 
classification if excess emission is detected (Section~\ref{sec:disks}).
If a source from 2MASS or WISE is resolved as a pair of candidates by Gaia, the
former is matched only to the component that is brighter in Gaia photometry.
118~Tau~A and B are not resolved in the AllWISE Source Catalog, but they are
resolved in the WISE All-Sky Source Catalog \citep{cut12a}, so the data from 
the latter have been adopted for the components of that system.

\section{Comparison to Previous Surveys}
\label{sec:prev}

\citet{rod14} proposed that the components of the wide binary system
LDS~5606 are members of the $\beta$~Pic moving group based on 
age diagnostics and estimates of $UVW$ velocities. 
However, they are among the candidate members of 32~Ori selected in
Section~\ref{sec:ident} and their $UVW$ velocities from 
Section~\ref{sec:uvw} agree closely with the bulk of the 32~Ori candidates.
In addition, the binary is located $\sim80$~pc from the center of 
the $\beta$~Pic group and $\sim30$~pc from the nearest candidate members from 
\citet{shk17}, but it resides within the volume inhabited by my 32~Ori 
candidates.

\citet{bel17} suggested that the eclipsing binary 2MASS J05525572$-$0044266
\citep{dra14} may be a member of 32~Ori based on the radial velocity of
the system. \citet{mur20} calculated a $UVW$ velocity for the binary
by combining a new measurement of the systemic radial velocity with
astrometry from Gaia DR2. Based on that velocity, they adopted the binary
as a member of 32~Ori, assigning it the designation of THOR~42.
However, \citet{mac21} found that the binary is older than the 32~Ori
association when the physical parameters of its components 
were interpreted with evolutionary models that include magnetic effects. 
That study also closely compared the $UVW$ data
for THOR~42 from \citet{mur20} and the velocities for other proposed members
of 32~Ori, concluding that the kinematics of THOR~42 may provide additional 
evidence of nonmembership. THOR~42 was not selected as a candidate member
of 32~Ori in Section~\ref{sec:ident} because it is slightly beyond the
declination boundary of my survey and it does not satisfy my kinematic
criteria. To examine its membership in more detail, I have calculated a new 
$UVW$ velocity using astrometry from Gaia DR3 and the radial velocity 
from \citet{mur20}, 
arriving at $(U,V,W)=(-15.9\pm0.4, -20.9\pm0.2, -5.4\pm0.1)$~km~s$^{-1}$.
Given its $XYZ$ position of ($-89.8, -45.4, -23.9$)~pc, the system's
$U$ and $V$ measurements are consistent with membership (although it would
be a spatial outlier in $Y$) but the $W$ velocity exhibits a fairly large
discrepancy ($\sim4$~km~s$^{-1}$) relative to the other candidates in my 
catalog (see Figure~\ref{fig:uvw}). Thus, this kinematic analysis supports the
suggestion from \citet{mac21} that THOR~42 is not a member of 32~Ori.

\citet{gag18d} used data from Gaia DR2 to search for new candidates for
members of several nearby associations, including 118~Tau and 32~Ori.
They identified nine candidates for either of the groups, eight of which
are among my candidates for 32~Ori. The one remaining candidate does not
satisfy the kinematic selection criteria from Section~\ref{sec:kin}.

\begin{figure}
\epsscale{1}
\plotone{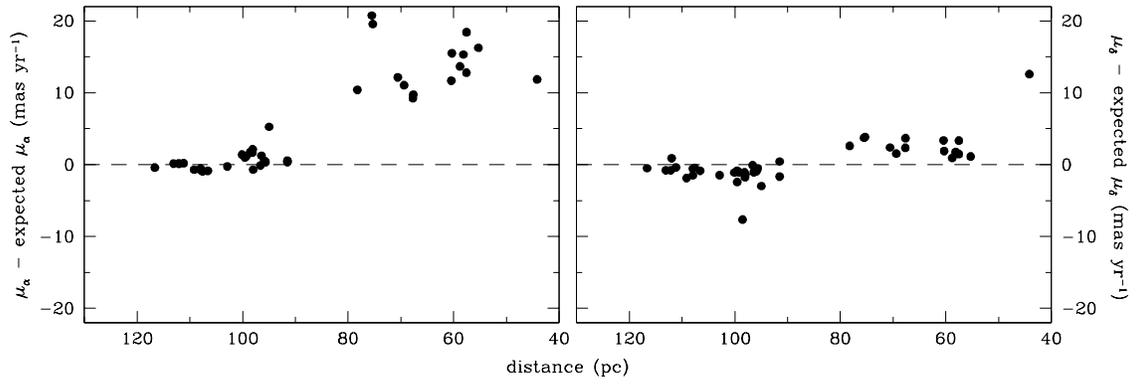}
\caption{
Proper motion offsets versus parallactic distance for candidate
members of 32~Ori-Columba from \citet{lee19} that have RUWE$<$1.6.
Only the candidates at $>$90~pc are among the candidate members of 32~Ori
selected in this work.
}
\label{fig:pp3}
\end{figure}

During analysis of $UVW$ velocities for previously identified members of
several nearby associations, \citet{lee19} proposed the existence of a single 
group that combines members of 32~Ori and a subset of members of the Columba 
association. Their new group contained 49 stars, 41 of which satisfy the 
criterion of RUWE$<$1.6 used in my kinematic selection of candidates 
(Section~\ref{sec:kin}). In Figure~\ref{fig:pp3}, I have plotted the proper
motion offsets and parallactic distances for those 41 stars. The 26 stars at 
distances of $>$90~pc correspond to some of the original members of 32~Ori 
while the 16 stars at $<$90~pc were originally identified as Columba members.
The data in Figure~\ref{fig:pp3} illustrate that those two subsets of stars
do not share similar kinematics, and hence do not represent a single
association.

2MASS J04435686+3723033 and 2MASS J04435750+3723031 likely comprise a wide
binary system \citep{schl10,phi20}.
Some studies have proposed that the system is a member of the $\beta$~Pic
moving group \citep{schl10,mal14,shk17} while others have found that its
membership in that association is uncertain \citep{mes17,phi20}.
Both components of the pair are among the candidate members of 32~Ori
selected in Section~\ref{sec:ident} and my calculations for their $UVW$
velocities in Section~\ref{sec:uvw} agree well with velocities of other
32~Ori candidates when the expansion pattern of the association is taken
into account, as shown in Figure~\ref{fig:uvw}.
For reference, the binary has an $XYZ$ position of ($-68.8, 17.6, -6.9$)~pc
and a $UVW$ velocity near $(-11, -19, -8)$~km~s$^{-1}$ (Table~\ref{tab:mem}). 
The system is located near the edge of the 32~Ori association in $XYZ$, 
but it is not an outlier.

\citet{sta20} proposed that HHJ~339 and HHJ~430 are members of the 32~Ori
association based on their evidence of youth and the available constraints
on their kinematics, primarily from Gaia DR2. Both stars are among the
candidates selected in Section~\ref{sec:ident}, so my analysis supports
the results from \citet{sta20}. The $U$ velocity of HHJ~339 is modestly 
discrepant, as shown in Figure~\ref{fig:uvw}, but it is retained in my
catalog of candidates (Section~\ref{sec:uvw}).

\citet{liu21} used data from Gaia DR2 to identify young groups within a volume
of space that overlaps with the volume considered in my survey for 32~Ori.
My sample of candidates for 32~Ori contains 32 stars that appear in group 11
from \citet{liu21}. That group includes one additional star (Gaia DR3 
3427527766435285248) that is absent from my catalog because it exceeds
my adopted threshold for RUWE.

\citet{ker21} used data from Gaia DR2 to identify groups of young stars
within a distance of a few hundred parsecs. Two of their groups, GT6 and GT7,
overlap with my catalog of candidates for 32~Ori.
For each group, \citet{ker21} presented a core sample and an expanded sample
in which contamination by non-members was expected to be higher.
The core/expanded samples for GT6 included 4/10 of the 11 members
of 118~Tau from \citet{mam16} while the core/expanded samples for GT7
contained 7/17 of the 44 members of 32~Ori from \citet{bel17} that have
Gaia DR3 parallaxes.
The total numbers of candidates in their core/expanded samples were
33/98 for GT6 and 11/25 for GT7; 31/50 of the stars in GT6 and all of the
stars in the two samples for GT7 appear in my catalog for 32~Ori.
Among the GT6 candidates from \citet{ker21} that are absent from my
sample of candidates, four are outside of the field that I have considered,
one satisfies all of my selection criteria except for RUWE, and the remaining
stars are rejected based on their photometry or kinematics.

\citet{kra17} suggested that the originally proposed members of 118~Tau
and 32~Ori might be kinematically related to the Taurus star-forming region.
Based on the kinematic data for my sample of 32~Ori candidates
(Section~\ref{sec:uvw}) and the kinematics of known members of Taurus 
\citep{luh18tau}, the median $UVW$ velocity of the 32~Ori association
differs by $\gtrsim5$~km~s$^{-1}$ from the median velocities of the
stellar aggregates in Taurus, which demonstrates that the 32~Ori candidates
did not originate from the Taurus clouds. Meanwhile, 
if the $XYZ$ positions of the 32~Ori association and the Taurus aggregates
are traced back in time using their median velocities (Section~\ref{sec:uvw2}),
32~Ori was never closer than $\sim15$~pc from any of the Taurus aggregates.
These kinematic and spatial separations between 32~Ori and Taurus are
common among young associations in the solar neighborhood 
\citep{bel17,gag18b}, so there is no evidence that 32~Ori and Taurus
have any meaningful relationship. The same has been found for other 
associations that are in the vicinity of Taurus but are kinematically
distinct from it \citep{luh18tau,gag20}.

\section{Properties of the 32 Ori Stellar Population}
\label{sec:pop}

\subsection{Initial Mass Function}
\label{sec:imf}

As in my previous studies of young clusters and associations, I have
used spectral type as an observational proxy for stellar mass when
characterizing the IMF of the 32~Ori association.
For 32~Ori candidates that lack spectral classifications and have
the necessary photometry, I have estimated spectral types from a comparison
of $G_{\rm BP}-G_{\rm RP}$, $G_{\rm RP}-J$, $J-H$, and $H-K_s$ to the
intrinsic values expected for young stars at various spectral types
\citep{luh22sc}. Three candidates lack measurements of
spectral types and colors, consisting of THOR~14Ab, THOR~9B, and HD~245924~B. 
Gaia DR3 provides a measurement of $G$ for the former but contains
no photometry for the latter two stars. I have adopted $K$-band data
for THOR~9B from \citet{bal09} and $V$-band data for HD~245924~B
from \citet{mas01}. The spectral types of those three stars were estimated by
combining their absolute magnitudes in those bands with the average relations 
between those magnitudes and spectral type for other 32~Ori candidates.

\begin{figure}
\epsscale{1.1}
\plotone{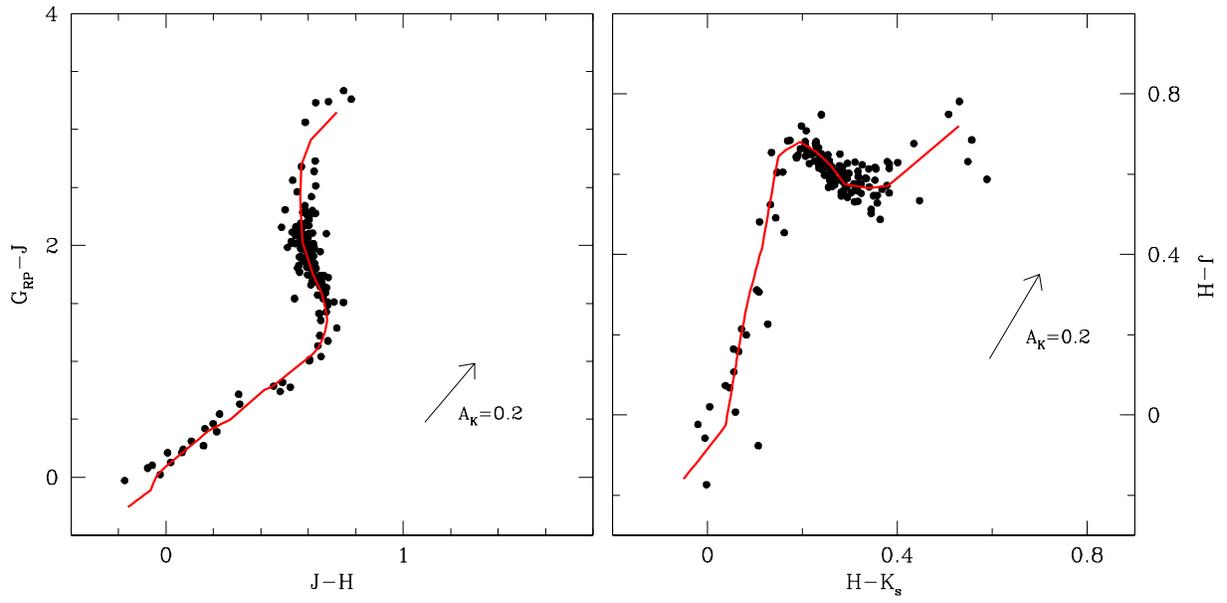}
\caption{
$G_{\rm RP}-J$ versus $J-H$ and $J-H$ versus $H-K_s$ for candidate members of
the 32~Ori association from Table~\ref{tab:mem}. The intrinsic colors of 
young stars from B0--M9 are indicated \citep[red lines,][]{luh22sc}.
}
\label{fig:cc}
\end{figure}

$G_{\rm RP}-J$, $J-H$, and $H-K_s$ for the 32~Ori candidates are plotted
in Figure~\ref{fig:cc} with the typical intrinsic colors of young stars.
Those data demonstrate that the candidates have little extinction 
\citep[$A_K\lesssim0.04$, see also][]{bel17}, so I have assumed that all 
candidates have no extinction when estimating their spectral types from colors. 
RX~J0437.4+1851~A and B and Gaia DR3 3241254244532020608, 
3416550830753546368, and 241638674007806336 have discrepant near-IR colors,
all of which are blended with companions or field stars. Those colors
are likely erroneous due to the blending, so they have been omitted from
Figure~\ref{fig:cc}.

\begin{figure}
\epsscale{1.2}
\plotone{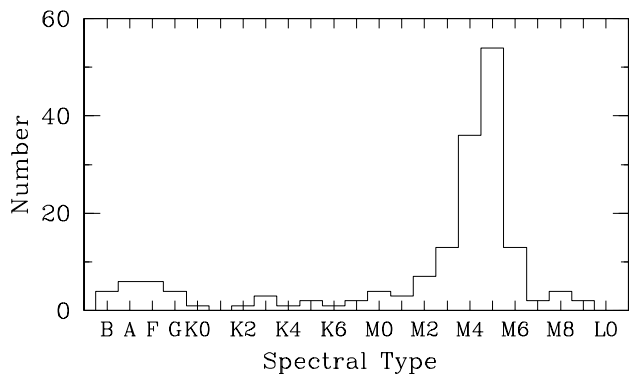}
\caption{
Histogram of spectral types for candidate members of the 32~Ori association
from Table~\ref{tab:mem}. For stars that lack spectroscopy, spectral types 
have been estimated from photometry (Figure~\ref{fig:cc}).}
\label{fig:histo}
\end{figure}

In Figure~\ref{fig:histo}, I have plotted a histogram of spectral types
for the 32~Ori candidates.
As discussed in Section~\ref{sec:comp}, this sample should have a high level
of completeness for spectral types earlier than $\sim$M7, which corresponds
to masses of $\gtrsim0.06$~$M_\odot$ for an age of $\sim20$~Myr 
\citep{bar98,bar15}.
The histogram exhibits a maximum near M5 ($\sim0.15$~$M_\odot$), and thus 
closely resembles the distributions measured for other young clusters
and associations in the solar neighborhood \citep[e.g.,][]{luh22sc}.

\subsection{Kinematic Ages}
\label{sec:uvw2}

Two kinds of kinematic ages are often investigated for unbound associations
of young stars, consisting of the expansion age and the traceback age
\citep{bla64,bro97,duc14,mam14,gol18,zar19,cru19,mir20,swi21}.
I have attempted to constrain these ages for the 32~Ori association
using the candidates in my catalog that have measurements of $UVW$ velocities.
The six modest kinematic outliers that are labeled in Figure~\ref{fig:uvw}
and discussed in Section~\ref{sec:uvw} are omitted from this analysis.

As discussed in Section~\ref{sec:uvw} and illustrated in Figure~\ref{fig:uvw},
the space velocities for the 32 Ori candidates exhibit correlations with
spatial positions that indicate the presence of expansion.
I have estimated the slopes of these correlations using robust
linear regression with bootstrap sampling, arriving at 
$0.050\pm0.014$, $0.027\pm0.006$, and $0.036\pm0.097$ km~s$^{-1}$~pc$^{-1}$
in $X$, $Y$, and $Z$, respectively.
The slope for $Z$ was derived only for $Z>-20$, which is the range in
which a correlation is evident in Figure~\ref{fig:uvw}.
The three slopes are not consistent with a single value, so the same
applies to the corresponding expansion ages, which span a range from
15--47~Myr at 1~$\sigma$.

\begin{figure}
\epsscale{1}
\plotone{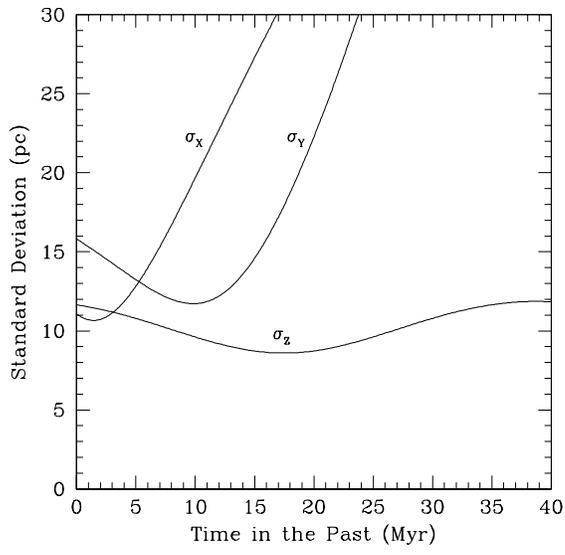}
\caption{
Standard deviations of $X$, $Y$, and $Z$ among candidate members of 32~Ori
over the last 40 Myr based on their current $UVW$ velocities and an epicyclic 
approximation of Galactic orbital motion.
}
\label{fig:trace}
\end{figure}

A traceback age can be estimated for an association by using the
kinematics of its members to identify the time in the past
when their spatial configuration was most compact.
For the 32~Ori candidates, I have calculated $XYZ$ positions as a function
of time over the last 40~Myr based on their measured $UVW$ velocities and
an epicyclic approximation of Galactic orbital motion \citep{mak04}.
Previous studies have considered various metrics to characterize the size
of an association for traceback analysis. I have adopted the standard
deviations of $X$, $Y$, and $Z$, which are plotted as a function of time
in the past in Figure~\ref{fig:trace}. Each axis exhibits a minimum during
the last 40~Myr, but the minima occur at significantly different times
that range from $\sim$1.5 to 17~Myr. As with the expansion ages, the
youngest traceback age is produced by the velocities along the $X$ axis.
The kinematic data for 32~Ori do not provide a well-defined 
age from either the expansion or the traceback of its candidate members,
which is a common result for young associations \citep{mam14,sod14}.

\subsection{Isochronal Ages}
\label{sec:iso}

\begin{figure}
\epsscale{1.1}
\plotone{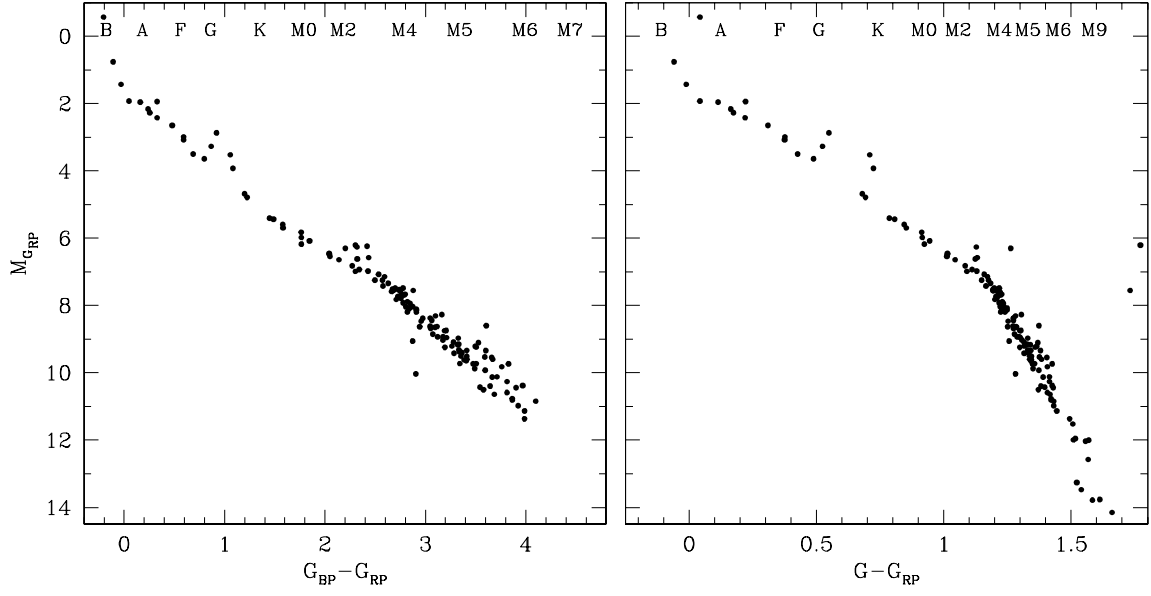}
\caption{
$M_{G_{\rm RP}}$ versus $G_{\rm BP}-G_{\rm RP}$ and $G-G_{\rm RP}$
for candidate members of the 32~Ori association from Table~\ref{tab:mem}.
The two stars that appear below the sequence in the left CMD are the
disk-bearing stars mentioned in Section~\ref{sec:phot}.
}
\label{fig:cmd2}
\end{figure}

By providing high-precision photometry and parallaxes, Gaia has enabled
the accurate measurement of the sequences of young associations in CMDs.
In Figure~\ref{fig:cmd2}, I have plotted $M_{G_{\rm RP}}$ versus
$G_{\rm BP}-G_{\rm RP}$ and $G-G_{\rm RP}$ for the candidate members of
32~Ori from Table~\ref{tab:mem}. The sequences in those CMDs are narrow
and well-defined, so they can be used to constrain the age of 32~Ori
relative to other associations. For that analysis, I consider stars
with $G_{\rm BP}-G_{\rm RP}=1.4$--2.8 (K5--M4), which should have
masses of $\sim$0.2--1~$M_\odot$ \citep{bar15}. 
Stars across that mass range are expected to exhibit similar evolution 
in their luminosities from $\sim1$--30~Myr \citep{her15}. 

\citet{luh20u} used data from Gaia DR2 to compare the relative ages of 
Sco-Cen populations and other young associations, including 32~Ori.
They found that the sequence of low-mass stars in 32~Ori \citep{bel17}
is $0.12\pm0.06$~mag brighter than that of the $\beta$~Pic moving group
\citep{bel15,gag18d}, which implied that the former is younger by $\sim3$~Myr
according to evolutionary models \citep{bar15,cho16,dot16,fei16}. 
\citet{bin16} estimated ages of 21/24$\pm4$~Myr for $\beta$~Pic from the
lithium depletion boundary using non-magnetic/magnetic models. UCL/LCC was
brighter than $\beta$~Pic by $0.10\pm0.02$~mag, indicating that 32~Ori
and UCL/LCC have similar ages.

\begin{figure}
\epsscale{1.4}
\plotone{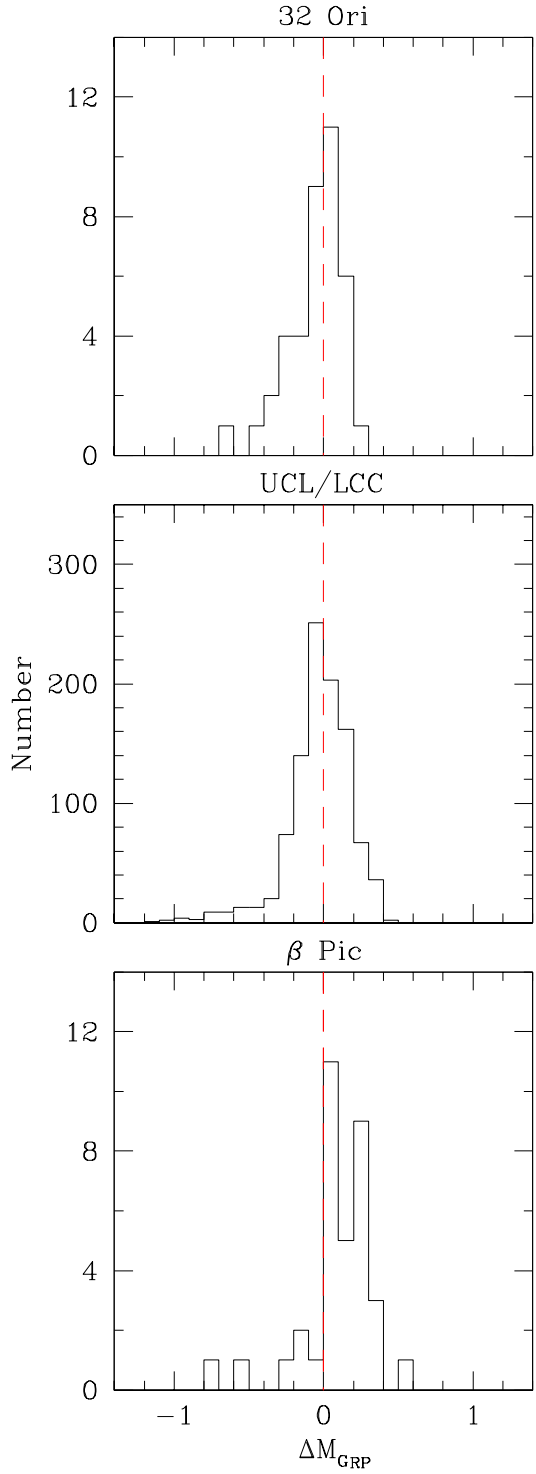}
\caption{
Histograms of offsets in $M_{G_{\rm RP}}$ from the median CMD sequence for
UCL/LCC for candidate members of 32~Ori, UCL/LCC, and $\beta$~Pic that have
$G_{\rm BP}-G_{\rm RP}=1.4$--2.8 ($\sim$0.2--1~$M_\odot$, K5--M4).
Negative values correspond to brighter magnitudes and younger ages.
}
\label{fig:ages}
\end{figure}

I have updated the previous comparison of 32~Ori, $\beta$~Pic, and UCL/LCC
to use data from Gaia DR3 and the new samples of candidates in 32~Ori
and UCL/LCC from Table~\ref{tab:mem} and \citet{luh22sc}, respectively.
For each association, I consider stars that have 
$G_{\rm BP}-G_{\rm RP}=1.4$--2.8, $\sigma_{\pi}<0.1$~mas, RUWE$<$1.6,
$\sigma_{BP}<0.1$, and $\sigma_{RP}<0.1$. Stars that have full disks are
excluded because accretion-related emission can contaminate the Gaia
colors \citep[Section~\ref{sec:disks},][]{luh22disks}.
The miscellaneous candidates in 32~Ori from Section~\ref{sec:add} are also
omitted since some of their parallactic distances may not be reliable.
The sample for UCL/LCC is defined with the criteria on kinematics and
celestial coordinates that were utilized by \citet{luh22disks}.
For each star in the three associations, I have calculated the offsets
in $M_{G_{\rm RP}}$ from the median sequences for UCL/LCC in
CMDs containing $G_{\rm BP}-G_{\rm RP}$ and $G-G_{\rm RP}$.
The mean offset from the two CMDs was adopted when $G$ was available.
Otherwise, only the offset from $M_{G_{\rm RP}}$ versus $G_{\rm BP}-G_{\rm RP}$
was used. Histograms of the resulting offsets are presented in
Figure~\ref{fig:ages}. The distributions for 32~Ori and UCL/LCC are
well-aligned with each other while the $\beta$~Pic sample is shifted to
fainter magnitudes, which is qualitatively consistent with the results
from \citet{luh20u}. For each pair of associations, I have calculated the
difference between their median $M_{G_{\rm RP}}$ offsets and I have
estimated the error in that difference using the median absolute deviations 
(MAD) for the distribution of differences produced by bootstrapping.
32~Ori and UCL/LCC are brighter than $\beta$~Pic by $0.13\pm0.06$~mag
and $0.12\pm0.04$~mag, respectively, and 32~Ori and UCL/LCC differ
by $0.00\pm0.04$~mag. These measurements are similar to the values from
\citet{luh20u}, so the age estimates from that study remain valid.
The difference of $0.00\pm0.04$~mag for 32~Ori and UCL/LCC
indicates that they are coeval to within $\pm1.2$~Myr according to
evolutionary evolutionary models.

As mentioned in Section~\ref{sec:prev}, groups GT6 and GT7 from
\citet{ker21} correspond to subsets of previously proposed members
of 118~Tau and 32~Ori, respectively.
That study reported significantly different ages of $16.8\pm2.0$ and
$27.2\pm3.8$~Myr for those groups, respectively.
To examine the relative ages of the subclusters, I have
selected two samples of 32~Ori candidates from the preceding age analysis,
one with $\alpha>75\arcdeg$ and $\delta<18\arcdeg$ (the 32~Ori subcluster)
and one with $\alpha>75\arcdeg$ and $\delta>18\arcdeg$ (the 118~Tau subcluster),
each of which contains 12 stars. 
The 118~Tau sample is brighter by $0.13\pm0.10$~mag, which should correspond
an age that is younger by $\sim3.6\pm2.9$~Myr based on evolutionary models. 
This age difference is smaller than that found by \citet{ker21} and 
has marginal significance.

\subsection{Circumstellar Disks}
\label{sec:disks}

I have used mid-IR photometry from WISE to check for evidence of
circumstellar disks around the candidate members of 32~Ori.
The images from WISE were obtained in bands centered at 3.4, 4.6, 12, and
22~$\mu$m, which are denoted as W1, W2, W3, and W4, respectively.
The angular resolution of Gaia DR3 is more than an order of magnitude higher
than that of the WISE images \citep{wri10,fab21}, so a close pair
of 32~Ori candidates can appear as a single unresolved source in WISE.
For each pair of this kind, the WISE source has been matched to the brighter
component in Gaia, as mentioned in Section~\ref{sec:uvw}.
For the 169 candidates in Table~\ref{tab:mem}, 
I have identified 160 matching sources in WISE.
I have visually inspected the AllWISE Atlas images of all of the WISE sources
to check for detections that are false or unreliable, which are
indicated by a flag in Table~\ref{tab:mem}. The measurements
for false detections have been omitted from Table~\ref{tab:mem} and
are not used in this work. 

\begin{figure}
\epsscale{1.2}
\plotone{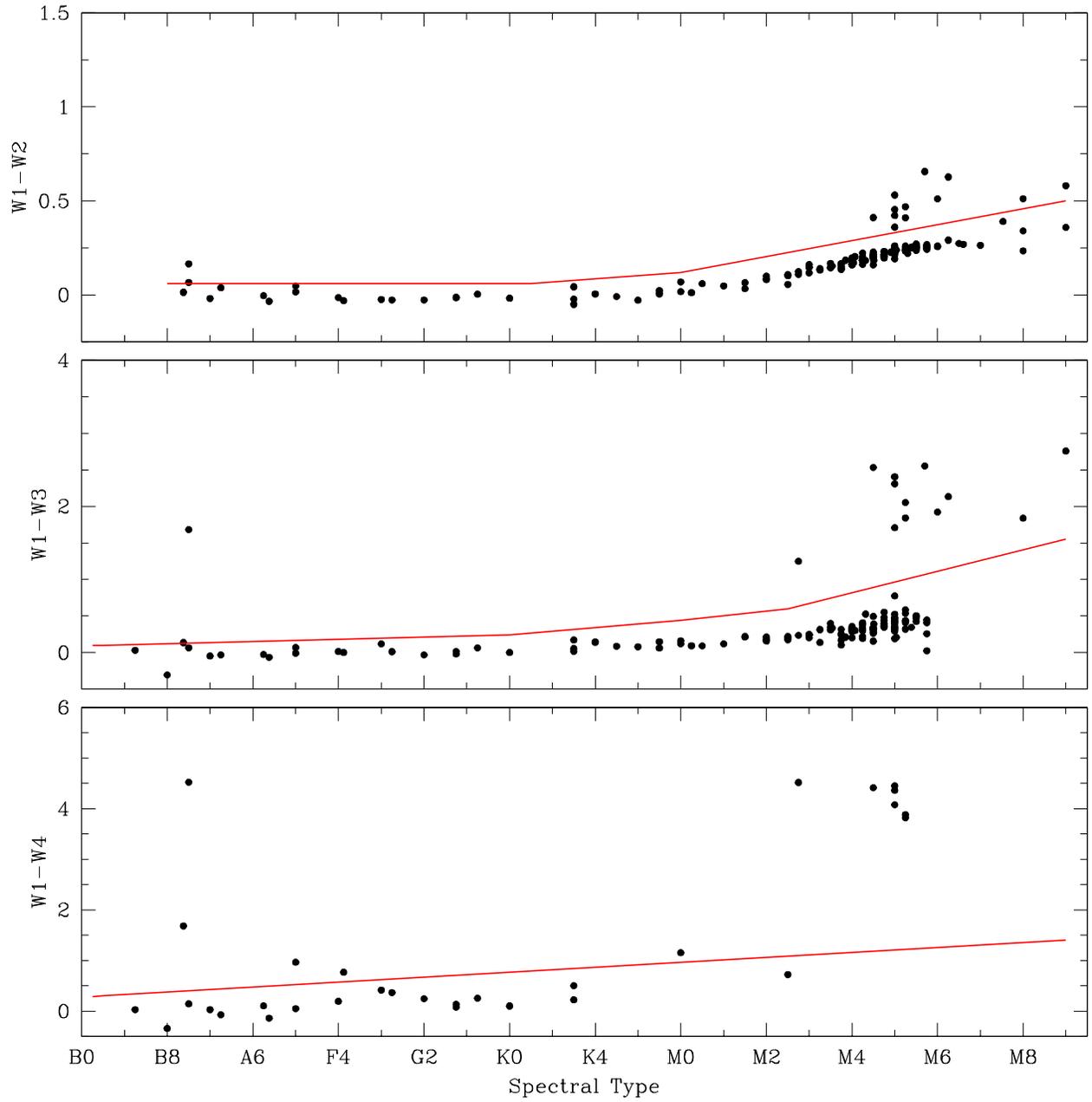}
\caption{
IR colors versus spectral type for candidate members of the 32~Ori
association from Table~\ref{tab:mem}.  For stars that lack spectroscopy,
spectral types have been estimated from photometry (Figure~\ref{fig:cc}).
In each diagram, the tight sequence of blue colors corresponds to stellar
photospheres. The thresholds used for identifying color excesses from
disks are indicated \citep[red solid lines,][]{luh22disks}.
}
\label{fig:exc1}
\end{figure}

As done in the disk survey by \citet{luh22sc}, I have used 
W1$-$W2, W1$-$W3, and W1$-$W4 to detect excess emission from disks.
In Figure~\ref{fig:exc1}, I have plotted those colors
versus spectral type for the WISE sources from Table~\ref{tab:mem}.
The W2 data at W2$<$6 have been omitted since they are subject to significant
systematic errors \citep{cut12b}.
For stars that lack spectroscopic measurements of spectral types, I have
adopted the photometric estimates from Section~\ref{sec:imf}.
In each of the three colors in Figure~\ref{fig:exc1}, most stars are found
in a well-defined sequence that corresponds to stellar photospheres.
A smaller number of stars are redder than that sequence,
indicating the presence of IR excess emission.
In each diagram in Figure~\ref{fig:exc1}, I have marked the threshold that was
used by \citet{luh22disks} for identifying color excesses.
If a star appears above a given threshold but a detection in any band
at a longer wavelength is consistent with a photosphere, an excess is
not assigned to the first band. In Table~\ref{tab:mem}, I have included
three flags that indicate whether excesses are identified in W2, W3, and W4.
Flags are absent for non-detections.

\begin{figure}
\epsscale{1.4}
\plotone{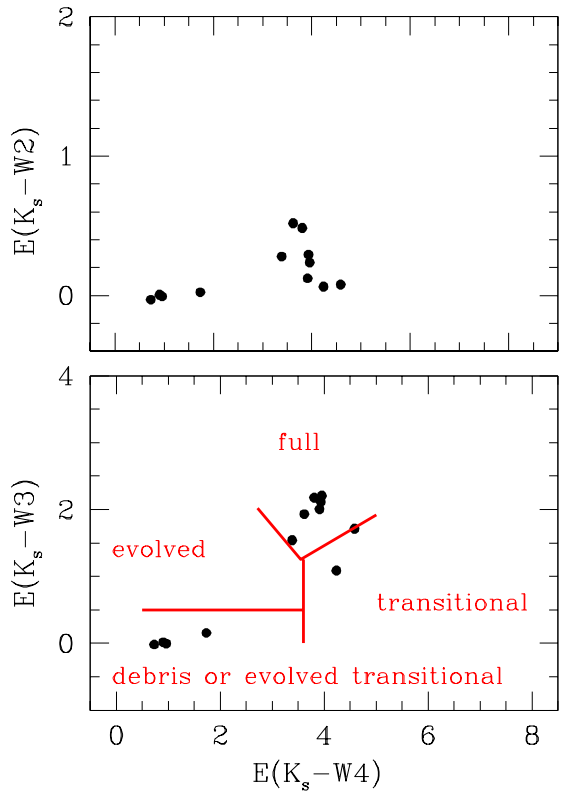}
\caption{
IR color excesses for candidate members of the 32~Ori association.
The boundaries used for assigning disk classes are shown in the bottom 
diagram (red solid lines).
}
\label{fig:exc2}
\end{figure}

For each source that has IR excess emission,
I have classified the evolutionary stage of its disk from among the following
options: full disk, transitional disk, evolved disk, evolved transitional 
disk, and debris disk \citep{ken05,rie05,her07,luh10,esp12}.
All of these classes except for the latter are considered primordial disks.
I have assigned these disk classes based on the sizes of the excesses in
$K_s-$W3 and $K_s-$W4 in the manner done in my previous disk surveys with WISE
\citep{luh12,esp14,esp18}.
The color excesses, E($K_s-$W3) and E($K_s-$W4), are calculated by
subtracting the expected photospheric color for a given spectral type
\citep{luh22sc}. The resulting excesses are plotted in Figure~\ref{fig:exc2}
with the criteria for the disk classes \citep{esp18}.  The excesses in $K_s-$W2 
have been included as well to illustrate the sizes of the excesses in W2.
As indicated in the diagram containing E($K_s-$W3) and E($K_s-$W4), the same
criteria are used for debris and evolved transitional disks, which are
indistinguishable in mid-IR data. Sources that lack excesses in any of
the WISE bands are omitted from Figure~\ref{fig:exc2} and are designated as
class~III \citep{lw84,lad87}. The sizes of the excesses for HD~36546 are 
indicative of a transitional disk according to my adopted classification
criteria, but I have labeled it as a debris disk based on more detailed
observations from previous studies \citep{mcd12,wu13,liu14,cur17,lis17}.

IR excesses from disks are detected for 18 of the 160 WISE sources.
Nine of those disks have been identified in previous work 
\citep{her08,mcd12,shv16,bel17,liu21}. The 18 disks consist of 12 full, 
one transitional, four debris or evolved transitional, and one debris.
The coolest disk-bearing sources are Gaia DR3 3320335175251540864 and 
3237866752286153088, which have spectral types of M8 and M9, corresponding
to masses of $\sim0.04$ and 0.02~$M_\odot$ \citep{luh03,bar15}.
\citet{bel17} reported W4 excesses for THOR~4B and THOR~38,
but I have classified those W4 detections as false or unreliable.

Among WISE sources in 32~Ori that have spectral types of $\leq$M6, the
fraction that have full, transitional, or evolved disks is
10/149=$0.07^{+0.03}_{-0.02}$. That value is consistent with the disk fraction 
for the same range of types in UCL/LCC 
\citep[$0.067^{+0.005}_{-0.004}$,][]{luh22disks},
which has a similar age as 32~Ori (Section~\ref{sec:iso}).

\section{Conclusions}

I have performed a survey for members of the 32~Ori association using
high-precision photometry and astrometry from Gaia DR3 and ground-based
spectroscopy. The new catalog of candidate members has been used to
characterize the IMF and age of the association and to identify and
classify its circumstellar disks. The results are summarized as follows:

\begin{enumerate}

\item
I have defined kinematic and photometric criteria for selecting candidate
members of 32~Ori using data from Gaia DR3 for previously proposed members
\citep{mam07,bel17} and stars associated with 118~Tau \citep{mam16}, which
are spatially adjacent to the 32~Ori members and share similar space motions.
I have applied the selection criteria to sources from Gaia DR3 that have 
$\alpha=50$--90$\arcdeg$, $\delta=0$--45$\arcdeg$, and distances of 70--120~pc.
The resulting sample contains 158 candidates, which includes previously
proposed members. I also have assigned candidacy to 17 additional stars from 
Gaia DR3 that do not satisfy the selection criteria but that have other 
evidence suggestive of membership (e.g., close companion to a candidate).

\item
I have used new and archival spectra to measure spectral types for 96
candidates, 62 of which have been classified for the first time in this work.
All of the age diagnostics in these spectra are consistent with the youth
expected for membership in 32~Ori, although some candidates lack the necessary
data for discrimination between young stars and field dwarfs. 
In addition, I have obtained high-resolution IR spectra of 39 candidates
to measure their radial velocities.

\item
I have calculated $UVW$ velocities for 113 of the 175 candidates that have
measurements of radial velocities.
In diagrams of $U/V/W$ versus $X/Y/Z$, most of the candidates are 
well-clustered and exhibit positive correlations, which indicates that 
they comprise a single association that is expanding. Six candidates are 
rejected based on discrepant $UVW$ velocities and other data, leaving 169 
candidates in the final catalog. The spatial distribution of the candidates 
shows two subclusters centered near the stars 32~Ori and 118~Tau, which is 
where most of the candidates from \citet{bel17} and \citet{mam16} are located.

\item
The new catalog of candidates for 32~Ori should have a high level of
completeness ($\sim90$\%) for spectral types of $\lesssim$M7 
($\gtrsim0.06$~$M_\odot$) and for the
ranges of celestial coordinates and distance that have been considered.
Based on the selection criteria in this survey, I estimate that my catalog
of candidates may include $\sim8$ field stars, likely consisting of a few
A/F/G stars and several M stars. Additional spectroscopy to measure
spectral types, age diagnostics, and radial velocities for candidates that
lack those data would help to reject such nonmenbers.

\item
The histogram of spectral types for the 32~Ori candidates exhibits a 
maximum at M5 ($\sim0.15$~$M_\odot$), indicating an IMF that has a similar
characteristic mass as other young clusters and associations in the solar
neighborhood.
 
\item
I have attempted to use the $UVW$ velocities that are available for
the 32~Ori candidates to derive kinematic ages for the association 
based on the expansion rates in $X$, $Y$, and $Z$ and the traceback
of $XYZ$ positions to their smallest configuration in the past.
Neither of these analyses produces a well-defined age.

\item
I have constrained the relative ages of the 32~Ori association,
the $\beta$~Pic moving group, and UCL/LCC in Sco-Cen using their
sequences of low-mass stars in CMDs constructed from Gaia data.
In terms of those sequences,
32~Ori and UCL/LCC are brighter than $\beta$~Pic by $0.13\pm0.06$~mag
and $0.12\pm0.04$~mag, respectively, and 32~Ori and UCL/LCC differ
by $0.00\pm0.04$~mag, indicating that 32~Ori and UCL/LCC are coeval to 
within $\pm1.2$~Myr and are younger than $\beta$~Pic by $\sim3$~Myr. 
These results are consistent with analysis of Gaia DR2 data for previous 
samples of candidates in 32~Ori and UCL/LCC \citet{luh20u}.
For reference, \citet{bin16} estimated ages of 21/24$\pm4$~Myr for
$\beta$~Pic from the lithium depletion boundary using
non-magnetic/magnetic models.

\item
I have used mid-IR photometry from WISE to identify 32~Ori candidates
that exhibit IR excesses from disks and I have classified the evolutionary
stages of the detected disks using the sizes of the excesses. 
Excesses are detected for 18 of the 160 WISE sources that are 
matched to the candidates from Gaia. Eight of these disks 
are reported for the first time in this work. 
Among candidates with spectral types of $\leq$M6, the fraction
that have full, transitional, or evolved disks is 
10/149=$0.07^{+0.03}_{-0.02}$, which is consistent with the value for UCL/LCC
\citep[$0.067^{+0.005}_{-0.004}$,][]{luh22disks}.

\end{enumerate}

\acknowledgements

I thank Bryson Cale, Peter Plavchan, and Adwin Boogert for assistance
with the analysis of the iSHELL data.
The IRTF is operated by the University of Hawaii under contract 80HQTR19D0030
with NASA. The observations at the CTIO 4~m Blanco telescope
were performed through program 2020B-0049 at NOIRLab.
This work used data provided by the Astro Data Archive at NOIRLab. 
CTIO and NOIRLab are operated by the Association of Universities for 
Research in Astronomy under a cooperative agreement with the NSF.
This work used data from the European Space Agency 
(ESA) mission Gaia (\url{https://www.cosmos.esa.int/gaia}), processed by
the Gaia Data Processing and Analysis Consortium (DPAC,
\url{https://www.cosmos.esa.int/web/gaia/dpac/consortium}). Funding
for the DPAC has been provided by national institutions, in particular
the institutions participating in the Gaia Multilateral Agreement.
2MASS is a joint project of the University of Massachusetts and IPAC
at Caltech, funded by NASA and the NSF.
WISE is a joint project of the University of California, Los Angeles,
and the JPL/Caltech, funded by NASA. This work used data from the 
NASA/IPAC Infrared Science Archive, operated by JPL under contract
with NASA, and the VizieR catalog access tool and the SIMBAD database, 
both operated at CDS, Strasbourg, France.
Guoshoujing Telescope (the Large Sky Area Multi-Object Fiber Spectroscopic Telescope LAMOST) is a National Major Scientific Project built by the Chinese Academy of Sciences. Funding for the project has been provided by the National Development and Reform Commission. LAMOST is operated and managed by the National Astronomical Observatories, Chinese Academy of Sciences.
Funding for the Sloan Digital Sky Survey IV has been provided by the
Alfred P. Sloan Foundation, the U.S. Department of Energy Office of
Science, and the Participating Institutions.
SDSS-IV acknowledges support and resources from the Center for High
Performance Computing  at the University of Utah. The SDSS
website is www.sdss.org.  SDSS-IV is managed by the
Astrophysical Research Consortium for the Participating Institutions
of the SDSS Collaboration including the Brazilian Participation Group,
the Carnegie Institution for Science, Carnegie Mellon University, Center for
Astrophysics | Harvard \& Smithsonian, the Chilean Participation
Group, the French Participation Group, Instituto de Astrof\'isica de
Canarias, The Johns Hopkins University, Kavli Institute for the
Physics and Mathematics of the Universe (IPMU) / University of
Tokyo, the Korean Participation Group, Lawrence Berkeley National Laboratory,
Leibniz Institut f\"ur Astrophysik Potsdam (AIP),  Max-Planck-Institut
f\"ur Astronomie (MPIA Heidelberg), Max-Planck-Institut f\"ur
Astrophysik (MPA Garching), Max-Planck-Institut f\"ur
Extraterrestrische Physik (MPE), National Astronomical Observatories of
China, New Mexico State University, New York University, University of
Notre Dame, Observat\'ario Nacional / MCTI, The Ohio State
University, Pennsylvania State University, Shanghai
Astronomical Observatory, United Kingdom Participation Group,
Universidad Nacional Aut\'onoma de M\'exico, University of Arizona,
University of Colorado Boulder, University of Oxford, University of
Portsmouth, University of Utah, University of Virginia, University
of Washington, University of Wisconsin, Vanderbilt University,
and Yale University.
The Center for Exoplanets and Habitable Worlds is supported by the
Pennsylvania State University, the Eberly College of Science, and the
Pennsylvania Space Grant Consortium.

\clearpage

\begin{deluxetable}{llll}
\tabletypesize{\scriptsize}
\tablewidth{0pt}
\tablecaption{Summary of Spectroscopic Observations\label{tab:log}}
\tablehead{
\colhead{Telescope/Instrument} &
\colhead{Mode/Aperture} &
\colhead{Wavelengths/Resolution} &
\colhead{Targets}}
\startdata
CTIO 4~m/COSMOS & red VPH/$1\farcs2$ slit & 0.55--0.95~\micron/4~\AA & 4 \\
CTIO 1.5~m/RC Spec & 47 Ib/$2\arcsec$ slit & 0.56--0.69~\micron/3~\AA & 1 \\
IRTF/SpeX & prism/$0\farcs8$ slit & 0.8--2.5~\micron/R=150 & 59 \\
IRTF/iSHELL & K$_{\rm gas}$/$0\farcs75$ slit & 2.17--2.46~\micron/R=49,000 & 39  \\
LAMOST & 540 l~mm$^{-1}$/$3\farcs3$ fiber & 0.37--0.9~\micron/5~\AA & 51
\enddata
\end{deluxetable}

\begin{deluxetable}{ll}
\tabletypesize{\scriptsize}
\tablewidth{0pt}
\tablecaption{Spectroscopic Data for Candidate Members of 32 Ori\label{tab:spec}}
\tablehead{
\colhead{Column Label} &
\colhead{Description}}
\startdata
Gaia & Gaia DR3 source name \\
RAdeg & Gaia DR3 right ascension (ICRS at Epoch 2016.0)\\
DEdeg & Gaia DR3 declination (ICRS at Epoch 2016.0)\\
SpType & Spectral type\tablenotemark{a}\\
Instrument & Instrument used for spectral classification\\
RVel & Barycentric radial velocity measured with iSHELL\\
e\_RVel & Error in RVel
\enddata
\tablenotetext{a}{Uncertainties are 0.25 and 0.5~subclass for optical and
IR spectral types, respectively, unless indicated otherwise.}
\tablecomments{
The table is available in its entirety in machine-readable form.}
\end{deluxetable}

\begin{deluxetable}{lllll}
\tabletypesize{\scriptsize}
\tablewidth{0pt}
\tablecaption{32~Ori Candidates Rejected by Space Velocities\label{tab:reject}}
\tablehead{
\colhead{Gaia DR3} &
\colhead{Other name} &
\colhead{$U$} &
\colhead{$V$} &
\colhead{$W$}\\
\colhead{} &
\colhead{} &
\multicolumn{3}{c}{(km~s$^{-1}$)}}
\startdata
3273765914308145792 & HD~24900 & $-30.1\pm0.1$ & $-20.7\pm0.1$ & $-22.3\pm0.1$ \\
225883054629988096 & HD~279481 & $-35.6\pm0.4$ & $-9.1\pm0.2$ & $-11.8\pm0.1$ \\
3387001696274562048 & \nodata & $-21.7\pm1.9$ & $-21.7\pm0.4$ & $-12.2\pm0.6$ \\
237427342611382016 & \nodata & $-6.5\pm0.2$ & $-20.3\pm0.1$ & $-7.5\pm0.1$ \\
3296359400790170880 & \nodata & $-22.0\pm0.2$ & $-20.0\pm0.1$ & $-11.52\pm0.02$ \\
164088748804295168 & HD~281691 & $-12.61\pm0.03$ & $-18.93\pm0.04$ & $-6.67\pm0.02$ 
\enddata
\end{deluxetable}

\begin{deluxetable}{ll}
\tabletypesize{\scriptsize}
\tablewidth{0pt}
\tablecaption{Candidate Members of 32 Ori from Gaia DR3\label{tab:mem}}
\tablehead{
\colhead{Column Label} &
\colhead{Description}}
\startdata
Gaia & Gaia DR3 source name \\
2MASS & 2MASS source name \\
WISEA & AllWISE source name \\
Name & Other source name \\
RAdeg & Gaia DR3 right ascension (ICRS at Epoch 2016.0)\\
DEdeg & Gaia DR3 declination (ICRS at Epoch 2016.0)\\
SpType & Spectral type \\
r\_SpType & Spectral type reference\tablenotemark{a} \\
Adopt & Adopted spectral type \\
pmRA & Gaia DR3 proper motion in right ascension\\
e\_pmRA & Error in pmRA \\
pmDec & Gaia DR3 proper motion in declination\\
e\_pmDec & Error in pmDec \\
plx & Gaia DR3 parallax\\
e\_plx & Error in plx \\
rmedgeo & Median of geometric distance posterior \citep{bai21}\\
rlogeo & 16th percentile of geometric distance posterior \citep{bai21}\\
rhigeo & 84th percentile of geometric distance posterior \citep{bai21}\\
RVel & Radial velocity \\
e\_RVel & Error in RVel \\
r\_RVel & Radial velocity reference\tablenotemark{b} \\
U & $U$ component of space velocity \\
e\_U & Error in U \\
V & $V$ component of space velocity \\
e\_V & Error in V \\
W & $W$ component of space velocity \\
e\_W & Error in W \\
Gmag & Gaia DR3 $G$ magnitude\\
e\_Gmag & Error in Gmag \\
GBPmag & Gaia DR3 $G_{\rm BP}$ magnitude\\
e\_GBPmag & Error in GBPmag \\
GRPmag & Gaia DR3 $G_{\rm RP}$ magnitude\\
e\_GRPmag & Error in GRPmag \\
RUWE & Gaia DR3 renormalized unit weight error\\
Jmag & 2MASS $J$ magnitude \\
e\_Jmag & Error in Jmag \\
Hmag & 2MASS $H$ magnitude \\
e\_Hmag & Error in Hmag \\
Ksmag & 2MASS $K_s$ magnitude \\
e\_Ksmag & Error in Ksmag \\
W1mag & WISE W1 magnitude \\
e\_W1mag & Error in W1mag \\
f\_W1mag & Flag on W1mag\tablenotemark{c} \\
W2mag & WISE W2 magnitude \\
e\_W2mag & Error in W2mag \\
f\_W2mag & Flag on W2mag\tablenotemark{c} \\
W3mag & WISE W3 magnitude \\
e\_W3mag & Error in W3mag \\
f\_W3mag & Flag on W3mag\tablenotemark{c} \\
W4mag & WISE W4 magnitude \\
e\_W4mag & Error in W4mag \\
f\_W4mag & Flag on W4mag\tablenotemark{c} \\
ExcW2 & Excess present in W2? \\
ExcW3 & Excess present in W3? \\
ExcW4 & Excess present in W4? \\
DiskType & Disk type
\enddata
\tablenotetext{a}{
(1) this work;
(2) \citet{abe14};
(3) \citet{bir20};
(4) \citet{opp97};
(5) \citet{can93};
(6) \citet{wic96};
(7) \citet{fin10};
(8) \citet{ngu12};
(9) \citet{kra17};
(10) \citet{schl12b};
(11) \citet{sle06};
(12) \citet{schl10};
(13) \citet{mal14};
(14) \citet{fou18};
(15) \citet{bow19};
(16) \citet{phi20};
(17) \citet{rod14};
(18) \citet{li98};
(19) \citet{esp19};
(20) \citet{bel17};
(21) \citet{her14};
(22) \citet{alc96};
(23) \citet{alc00};
(24) \citet{bia12};
(25) \citet{liu21};
(26) \citet{gag15c};
(27) \citet{hou99};
(28) \citet{cow69};
(29) \citet{abt95};
(30) \citet{gre99};
(31) \citet{abt08};
(32) \citet{sua17};
(33) \citet{bri19};
(34) \citet{abt77};
(35) \citet{edw76};
(36) \citet{bin15};
(37) \citet{abt04}.}
\tablenotetext{b}{
(1) Gaia DR3;
(2) \citet{abd22};
(3) \citet{ngu12};
(4) \citet{zha21} and LAMOST DR7;
(5) this work;
(6) \citet{kra17};
(7) \citet{fou18};
(8) \citet{phi20};
(9) \citet{rod14};
(10) \citet{bel17};
(11) \citet{zun21};
(12) \citet{alc00};
(13) \citet{gon06};
(14) \citet{gre99};
(15) \citet{whi07};
(16) \citet{sou18}.}
\tablenotetext{c}{nodet = nondetection; false = detection from
AllWISE appears to be false or unreliable based on visual inspection.}
\tablecomments{
The table is available in its entirety in machine-readable form.}
\end{deluxetable}

\clearpage

\end{document}